\documentclass{nature}

\usepackage{verbatim}
\newcounter{mybibstartvalue}

\usepackage{amsmath,amssymb}
\usepackage{graphicx}
\usepackage{caption}
\usepackage{lineno}

\makeatletter
\let\saved@includegraphics\includegraphics
\AtBeginDocument{\let\includegraphics\saved@includegraphics}
\renewenvironment*{figure}{\@float{figure}}{\end@float}
\makeatother

\title{Shock cooling of a red-supergiant supernova at redshift 3 in lensed images}

\author{Wenlei Chen*$^1$, Patrick L. Kelly$^1$, Masamune Oguri$^{2,3,4}$, Thomas J. Broadhurst$^{5,6,7}$, Jose M. Diego$^8$, Najmeh Emami$^1$, Alexei V. Filippenko$^9$, Tommaso L. Treu$^{10}$, and Adi Zitrin$^{11}$}

\begin{document}



\maketitle

\begin{affiliations}
 \item School of Physics and Astronomy, University of Minnesota, 116 Church Street SE, Minneapolis, MN 55455, USA
 \item Center for Frontier Science, Chiba University, 1-33 Yayoi-cho, Inage-ku, Chiba 263-8522, Japan
 \item Research Center for the Early Universe, University of Tokyo, Tokyo 113-0033, Japan
 \item Kavli Institute for the Physics and Mathematics of the Universe (Kavli IPMU, WPI), University of Tokyo, 5-1-5 Kashiwanoha, Kashiwa, Chiba 277-8583, Japan
 \item University of the Basque Country UPV/EHU, Department of Theoretical Physics, Bilbao, E-48080, Spain
 \item DIPC, Basque Country UPV/EHU,San Sebastian, E-48080, Spain
 \item Ikerbasque, Basque Foundation for Science, Bilbao, E-48011, Spain
 \item IFCA, Instituto de F\'isica de Cantabria (UC-CSIC), Av. de Los Castros s/n, 39005 Santander, Spain
 \item Department of Astronomy, University of California, Berkeley, CA 94720-3411, USA
 \item Department of Physics and Astronomy, University of California, Los Angeles, CA 90095, USA
 \item Physics Department, Ben-Gurion University of the Negev, P.O. Box 653, Beer-Sheva 8410501, Israel
\end{affiliations}


\begin{abstract}
The core-collapse supernova of a massive star rapidly brightens when a shock, produced following the collapse of its core, reaches the stellar surface. As the shock-heated star subsequently expands and cools, its early-time light curve should have a simple dependence on the progenitor's size\cite{waxman17} and therefore final evolutionary state.
Measurements of the progenitor's radius from early light curves exist for only a small sample of nearby supernovae\cite{valenti2014,garnavich2016,sn2011dh-1,sn2011fu,ptf12gzk,sn2016gkg,tartaglia2017,arcavi2017,p21,szalai2019,rui2019,xiang2019,sn2018fif}, and almost all lack constraining ultraviolet observations within a day of explosion.
The several-day time delays and magnifying ability of galaxy-scale gravitational lenses, however, should provide a powerful tool for measuring the early light curves of distant supernovae, and thereby studying massive stellar populations at high redshift. Here we analyse individual rest-frame ultraviolet-through-optical exposures taken with the {\it Hubble Space Telescope} that simultaneously capture, in three separate gravitationally lensed images, the early phases of a supernova at redshift $z \approx 3$  beginning within $5.8\pm 3.1$\,hr of explosion. The supernova, seen at a lookback time of $\sim11.5$ billion years, is strongly lensed by an early-type galaxy in the Abell~370 cluster. We constrain the  pre-explosion radius to be $533^{+154}_{-119}$ solar radii, consistent with a red supergiant.
Highly confined and massive circumstellar material at the same radius can also reproduce the light curve, but is unlikely since no similar low-redshift examples are known.
\end{abstract}


As part of a search for gravitationally lensed, transient events in archival {\it Hubble Space Telescope (HST)} imaging of the {\it Hubble} Frontier Fields (HFF)\cite{lotz2017}, we discovered three images of a rapidly evolving, strongly lensed supernova (SN) in imaging of the Abell~370 galaxy cluster ($z=0.375$\cite{strublerood1991}) field obtained in 2010 December. 
Fig.\,1 shows these sources, which we label S1, S2, and S3 (as shown in Fig.\,1d), and images of the underlying strongly lensed host galaxy, which is $\sim1'$ away from the galaxy cluster's center.

As shown in Extended Data Figs.\,2g and 2h, we measure the photometric redshift of the images of the SN's host galaxy, and find a joint probability distribution of $z=2.93^{+0.06}_{-0.05}$ using {\tt BPZ}\cite{bpz1} and $z=2.94^{+0.06}_{-0.07}$ using {\tt EAZY}\cite{eazy}. 
The images' fluxes are listed in Extended Data Table\,1a.

We construct a lens model of the early-type galaxy and galaxy cluster using {\tt GLAFIC}\cite{kawamataishigakishimasaku18,kawamataoguriishigaki16,oguri10,oguri21}, which predicts that the early-type galaxy creates a total of four magnified images of the SN host galaxy in an Einstein-cross configuration, as shown in Fig.\,1e. One image, however, is insufficiently magnified to detect given the underlying galaxy.
As listed in Extended Data Table\,1b, the model predicts that image S1 has a time delay of $30.6 \pm 5.6$\,days relative to S3 and of $9.6 \pm 2.3$\,days relative to S2, while S2 has a time delay of $21.0 \pm 3.4$\,days relative to S3. 
Taking into account the factor of 4 redshift time dilation of sources at $z = 3$, the SN as seen in image S1 is $7.7 \pm 1.4$\,days younger than it appears in image S3, while image S2 shows the SN as it was $5.3 \pm 0.9$\,days younger than it appears in image S3. According to the lens model, images S1, S2, S3, and S4 have respective magnifications of $4.2 \pm 1.1$, $7.8 \pm 1.3$, $5.8 \pm 0.8$, and $1.2 \pm 0.5$. 

We next fit separate blackbody models to the spectral energy distributions (SEDs) of the three delayed images of the SN. Using the predicted magnification values from the lens model, we find effective blackbody temperatures of $5.5_{-3.0}^{+13.8}\times10^{4}$\,K, $2.0_{-0.2}^{+0.2}\times10^{4}$\,K, and $1.1_{-0.1}^{+0.1}\times10^{4}$\,K for SN images S1, S2, and S3, respectively.
We show, in Fig.\,1e, a pseudocolour image of the SN constructed by assigning the F160W, F110W, and F814W difference images to the red, green, and blue (RGB) channels. 
Fig.\,1e shows the rapid change in the SN's colour as it cools from
$\sim 100,000$\,K to $\sim 10,000$\,K over $\sim8$\,days in the rest frame at $z=3$.
As described in Methods, we find that the optical depth for microlensing is very low, precluding chromatic microlensing.

Following shock breakout, the photosphere rapidly cools as the thermal energy of the material is converted to radiation and the kinetic energy of the expanding envelope.
Emission escapes when the optical depth of the shock drops below $c/v$, where $c$ and $v$ are (respectively) the speed of light and the shock velocity\cite{ns10,rw11}.
Therefore, a shell of circumstellar material (CSM) can dominate the emission associated with shock cooling only if its optical depth before explosion exceeds $c/v$\cite{np14,p15}.

To calculate the predicted light curves of the core-collapse SNe of supergiant stars where the CSM does not dominate the shock-cooling emission, we adopt an analytical model \cite{sw17} with a planar-phase correction\cite{sn2018fif}. We label these models blue supergiant (BSG) and red supergiant (RSG), respectively.
Modelling of the light curves of nearby Type IIP SNe shows that a significant fraction of progenitors have CSM at radii $>1000$\,$R_{\odot}$\cite{morozova2017,morozova2018}, although almost all low-redshift datasets lack ultraviolet (UV) photometry within a day of explosion, which would provide additional constraints on the radius of optically thick material as we describe in the Methods.
We therefore consider a pair of analytical models of a SN explosion surrounded by CSM which we refer to as CSM-homologous\cite{p21} and CSM-planar\cite{m21}.

Although our ten fitting parameters exceed the nine measured data points, the four parameters associated with the lens model are constrained by informative priors. We are able to use the Bayes factor to distinguish among potential models and marginalise fully over all parameters.
We compute the Bayes factor for each SN model relative to the null hypothesis that the lensed source is a blackbody whose temperature and luminosity do not evolve with time. 
In Extended Data Table\,1c, we list the natural logarithm of the ratio of the Bayes factors, and the ratios provide strong evidence against the null hypothesis. 
Given two hypothetical models A and B, $\log B_{AB}>2$ is considered to be very strong evidence in favour of model A, while $\log B_{AB}<0.5$ indicates little or no statistical evidence\cite{kass1995}.
When we assume that the CSM has a mass within a range of 0.001--0.1\,$M_\odot$ that is consistent with those used by modelling efforts\cite{p21,m21}), the RSG model is significantly favoured in comparison to both the BSG and CSM models (with $\Delta \log B_{\rm AB} > 2$).
In Fig.\,2, we plot our constraints on the progenitor radius, envelope mass, and shock velocity for the RSG model, and we plot the posterior probability distributions for all model parameters in Extended Data Fig.\,4.
The best-fit parameter values and their confidence intervals for all models are listed in Extended Data Table\,2.

If we allow the CSM mass to become as large as $1\,M_\odot$, the Bayes factor for the CSM-homologous model becomes
better than (but not significantly, with $\Delta \log B_{\rm AB} = 0.7$) that of the RSG model, while the difference between the RSG model and the Bayes factor of the CSM-planar CSM model decreases. However, the CSM-homologous model yields a CSM radius of $481^{+157}_{-118}\,R_{\odot}$, and prefers a CSM mass of 1\,$M_{\odot}$. The CSM radius is smaller than the $>1000$\,$R_{\odot}$ inferred for a sample of SN~IIP explosions of RSGs\cite{morozova2018}, and the dense, compact CSM-homologous model approximates the outer atmosphere of the RSG model. 

The explosion of an RSG, the most common SN type in the local universe \cite{li2011}, is the most probable explanation for the observations. As shown in Extended Data Fig.\,7, the early light curve and the evolution of the temperature of this event are similar to those of the SN~IIP explosions of RSG progenitor stars. Extended Data Fig.\,7 shows that the observations are not well matched by those of well-studied SNe~IIP whose progenitors were inferred, from their light curves, to have CSM. Likewise, early available UV observations of SNe~Ia, Ib/c, IIb, or FBOTs are not well matched to our photometry.

In Fig.\,3, we show two illustrative examples of the reconstructed light curve of the SN. These plot the light curve when the images' magnification and times delays are set to their best-fitting values, and after they are shifted by $1\sigma$ (standard deviation) to smaller values; see Extended Data Table\,1b for lens values). For each light curve, we show the best-fitting RSG model light curve, and the corresponding constraints on its pre-explosion radius, which are $615^{+86}_{-76}\,R_{\odot}$ and $509^{+74}_{-67}\,R_{\odot}$, respectively. These demonstrate the dependence of the inference about the progenitor's size on the lens-model uncertainties. The constraints in Fig.\,3 are consistent with but shifted by a small amount from our primary estimate, since the SN model and photometry together favour larger magnifications (by $\lesssim 1.2\sigma$) and smaller time delays (by $\lesssim 0.5\sigma$).
In the rest frame of the SN at $z \approx 3$, the central wavelength of the ACS-WFC F814W (broad $I$) filter corresponds to 
2000\,\AA, 
WFC3-IR F110W (wide $yJ$) to 
2900\,\AA, 
and WFC3-IR ($H$) to
3900\,\AA.
Fig.\,3 shows that the luminosity measured in the F814W filter grew and then faded, while that in the reddest (F160W) filter only increased.

Discoveries of strongly gravitationally lensed SNe\cite{kellyrodneytreu15,goobar2017,rodney2021} offer a new avenue for probing the populations of massive stars present in the high-redshift universe\cite{oguri2019,foxleymarrable2020}. Two multiply imaged SNe (SN Refsdal\cite{kellyrodneytreu15} and this work) have been discovered in the six HFFs, and we use these to estimate the volumetric core-collapse SN (CCSN) rate to $z=3$. 
Using our simulations of the CCSN light curves, we constrain the volumetric CCSN rate in four redshift bins from 0.5 to 6. 
Given these two magnified SNe, we obtain
CCSN rates of $4.1_{-0.8}^{+17.0}\times 10^{-4}\,\mathrm{yr}^{-1}\,\mathrm{Mpc}^{-3}$ at $z=1$--2 and $7.6_{-1.3}^{+31.8}\times 10^{-4}\,\mathrm{yr}^{-1}\,\mathrm{Mpc}^{-3}$ at $z=2$--3.5, as shown in Fig.\,4. 

The principal existing constraints on the CCSN rates to $z \approx 2.5$ are from an analysis of the Cosmic Assembly Near-infrared Deep Extragalactic Legacy Survey (CANDELS)\cite{candels} and the Cluster Lensing And Supernova survey with {\it Hubble} (CLASH)\cite{clash} (the CANDELS+CLASH analysis)\cite{strolger15}. For a given survey, the expected number of CCSNe depends on both the intrinsic volumetric rate of CCSNe and an assumed prior on extinction due to dust. For a given number of detected CCSNe, greater average extinction will lead a greater inferred volumetric CCSN rate, since an increased fraction SNe will be too faint to detect. To draw comparison with the CANDELS+CLASH constraints, we must use both the same inferred volumetric CCSN rates and the extinction prior 
to predict the expected number of strongly lensed SNe in archival observations of the HFF cluster fields. We find a probability of only $p=0.026$ of discovering at least two strongly lensed CCSN. 

The exponential scale factor of $0.187$\,mag used by the CANDELS+CLASH analysis\cite{strolger15} to construct a dust-extinction prior corresponds to an extreme average $A_V = 5.35$\,mag which differs from the average of the sample ($A_V = 0.405$) used to construct it\cite{schmidt1994,hamuy2002}. For our primary analysis, we instead use a prior with mean $A_V = 1$\,mag informed by measurements of low-redshift SNe\cite{kelly2012,drout2011,prentice2016}, as shown in Extended Data Fig.\,8a.

The rate of CCSNe per unit volume ($R_\mathrm{CC}(z)$) can be expressed as a function of the volumetric star-formation density $\psi(z)$: $R_\mathrm{CC}(z) = \psi(z)\, k_\mathrm{CC}$.  Given the redshift-dependent $\psi(z)$ inferred from observations of galaxies\cite{MD14}, and the two multiply imaged SN in the HFF fields, we obtain $k_\mathrm{CC}=0.0087_{-0.0046}^{+0.0071}\,M_\odot^{-1}$ and a maximum-likelihood estimate of $k_\mathrm{CC}=0.0060\,M_\odot^{-1}$. In contrast to previous analysis of high-redshift CCSNe rates\cite{strolger15}, these values are consistent with the theoretical expectation of 
$k_\mathrm{CC}=0.0068\,M_\odot^{-1}$ for a Salpeter initial-mass function (IMF)\cite{salpeter_imf} and the approximate assumption that stars with initial masses between 8 and 40\,$M_{\odot}$ successfully explode\cite{MD14}.


\begin{methods}

\subsection{Cosmology:}

In this paper, we assume a concordance cosmology described by the $\Lambda$ cold dark matter model with $\Omega_m=0.3$, $\Omega_\Lambda=0.7$, and a Hubble constant H$_0=70\,\mathrm{km}\,\mathrm{s}^{-1}\,\mathrm{Mpc}^{-1}$.

\subsection{HST imaging and transient searching:}

All {\it HST} images are aligned using the {\tt TweakReg} tool and resampled to a scale of $0.03 ''$\,pixel$^{-1}$ using the {\tt AstroDrizzle} package\cite{fruchter10}. We prepared our template images using the full set of the {\it HST} imaging from the WFC3 and ACS cameras.  We grouped and coadded images from each {\it HST} visit and then created a difference image by subtracting a deep template image from the coaddition. We next identified all peaks in the resulting difference images with flux exceeding a $3\sigma$ signal-to-noise ratio, and classified the peaks by a convolutional neural network (CNN) machine specifically developed for the transient search of the archival {\it HST} imaging. For photometry in the difference imaging, we used {\tt PythonPhot}\cite{jonesscolnicrodney15}, which includes an implementation of point-spread-function (PSF) fitting photometry based on the {\tt DAOPHOT} algorithm\cite{stetson87}.
Our search led us to identify this SN from {\it HST} images, as well as the already known SN Refsdal in the MACS J1149.5+2223 galaxy-cluster field\cite{kellyrodneytreu15}.
The SN images were detected in the {\it HST} imaging taken in 2010 December. Prior to the 2010 December {\it HST} visit, the field was observed most recently in 2009 September. The field was not observed again until 2014 October, and neither set of images
from observations before 2009 and after 2014  
show evidence of variability. 
The relative time delays and magnifications of the three strongly lensed images allow us to reconstruct the light curve of the SN beginning within a day after explosion, which we use to constrain the radius of its progenitor.

\subsection{Photometry of the SN images:}

A difference image is constructed from a pair of coadditions of images acquired at two epochs. However, {\it HST}'s roll angle generally changes between visits. Consequently, the four diffraction spikes of the PSF will rotate, and a simple subtraction of the imaging contains a negative and positive set of residual spikes. In the case of the lens galaxy in the centre of the multiply imaged SN, as shown in Extended Data Fig.\,1d, these residual spikes coincide with the SN images. To remove these spikes, we created a model of the lens galaxy and subtracted it from each set of images, after convolving the model with the PSF measured from the imaging acquired at each epoch.

We used {\tt GALFIT}\cite{galfit} to model the lens galaxy and adjacent sources in deep imaging taken by the HFF survey. Extended Data Fig.\,1a--1c show the {\it GALFIT} models. After convolution with the PSF measured for each epoch, we subtracted the best-fit {\tt GALFIT} models from both the HFF coaddition, and the previous epoch of imaging.
This method yielded much improved contamination of the difference images, as shown in Extended Data Fig.\,1e. 
To generate a deep template for WFC3-IR F110W, we rescaled the WFC3-IR F105W and F125W templates to have zeropoints that match that of the WFC3-IR F110W imaging.

\subsection{Fourth image of SN host galaxy:}

We are not able to detect G4 in coadded images, as expected given its small magnification and underlying early-type galaxy. According to the predicted magnification ratio of G2 to G4 and the photometry of G2, the apparent magnitude of G4 is expected to be 28.1~AB in the ACS-WFC F814W band and 27.7~AB in the WFC-IR F160W band, fainter than the $3\sigma$ (standard deviation) noise levels, where the $3\sigma$ noise levels within an aperture for G2 are 28.0~AB in the ACS-WFC F814W band and 27.4~AB in the WFC-IR F160W band. The appearance of the fourth image of the SN (S4) should be the most delayed. As we will describe, we find that the SN exploded only $\sim 1$\,day before we observed it in image S1. The lens model predicts that, in individual visits to the field, we see G4 as it was $\sim23$\,days before we see G1. Consequently, the SN should not yet be visible in G4 when the 2010 data were taken, and we do not detect it.

\subsection{Evaluating whether the SN images' colour differences could arise from noise:}
We sought to ascertain whether the observed colour differences among images S1--S3 could be explained by flux-measurement uncertainties, given a lensed transient with an unchanging colour.
As our null hypothesis, we consider a blackbody source with a constant temperature and luminosity, such that differences in flux arise only from  magnification and uncertainties associated with the flux measurement. The alternative hypothesis is that the SN has distinct effective temperatures and luminosities across the three observed images. We placed a broad prior on the luminosity of the blackbody between the solar value ($3.828\times10^{26}$\,W) and that of  ASASSN-15lh\cite{dong2016} ($2.2\times10^{38}$\,W).
The likelihood-ratio test statistic $-2\ln{\Lambda}\approx43$, where $\Lambda$ is the ratio between likelihoods of the null and alternative hypotheses.
This test statistic asymptotically approaches the chi-squared ($\chi^{2}$) distribution with degree of freedom equal to 4 (because the alternative hypothesis has four more free parameters than the null hypothesis), according to Wilks' theorem. Hence, the likelihood-ratio test gives a $p$-value of $\sim1.0\times10^{-8}$, equivalent to a $\sim5.6\sigma$ significance for a normal-distributed variable.
We constrained the effective temperature ($T$) and luminosity ($L$) using a Markov Chain Monte Carlo (MCMC) analysis and uniform priors for $10^3\,\mathrm{K}<T<10^9\,\mathrm{K}$ and $3.828\times10^{26}\,\mathrm{W}<L<2.2\times10^{38}\,\mathrm{W}$. The  posterior distributions are shown in Extended Data Fig.\,1f--1h.

\subsection{Photometric redshift:}

We conducted photometry of the multiple galaxy images G1, G2, and G3 from the published coaddition of the {\it HST} imaging collected from the HFF survey\cite{lotz2017}. 
Before the measurement, we used {\tt GALFIT} to fit the bright galaxy in the centre of the system and subtracted the best-fit model from the field. We then fit photometry using {\tt BPZ}\cite{bpz1} and {\tt EAZY}\cite{eazy} software to estimate the photometric redshift of the multiply imaged galaxy. As shown in Extended Data Figs.\,2g and 2h, both software packages yield a photometric redshift consistent with $z\approx3$. 
We performed a likelihood-ratio test for a hypothesis assuming three independent redshifts against the hypothesis of a multiply imaged galaxy. The test statistic $2\ln{\Lambda}\approx3.5$, providing only a $\sim0.9\sigma$ significance in favour of the hypothesis for unrelated galaxy images with independent redshifts. This indicates that the disagreement among the photometric-redshift results from the three images is statistically insignificant. The 95\% confidence interval of the photometric redshift from the joint probability is $(2.85,~3.05)$ from {\tt BPZ} and $(2.81,~3.07)$ from {\tt EAZY}. This result is also consistent with the photometric redshift from a previous study\cite{bradacmaruhuang19}, in which the 95\% confidence intervals of the G1, G2, and G3 redshifts are $(3.122,~3.513)$, $(2.888,~3.351)$, and $(2.736,~3.298)$, respectively.
The next most probable value of the photometric redshift is $z \approx 0.2$, but we can exclude this redshift range because it would lie in the foreground of the  Abell 370 cluster at redshift $z=0.375$.
Fitting stellar population synthesis models\cite{m11_stellarmodel} to the host galaxy's SED, we measure a stellar mass of $\sim 2\times10^8\,M_\odot$ and a star-formation rate of $\sim 0.2\,M_\odot\,\mathrm{yr}^{-1}$. Extended Data Fig.\,2i shows the distribution of CCSN host-galaxy stellar mass and star-formation rate\cite{schulze2021}, where the red star marks the host galaxy of the newly discovered SN.

\subsection{Spectroscopic observations of the SN host galaxy:}

We acquired optical spectra of the host galaxy using the Blue Channel Spectrograph\cite{mmtblue}
on the MMT telescope (MMT-Blue;
6 Oct. 2019 UT and 23 Dec. 2019 UT), as well as near-infrared spectra using the Large Binocular Telescope Utility Camera in the Infrared\cite{luci} (LBT-LUCI;
30 Oct. 2021 UT) and the Multi-Object Spectrometer For Infra-Red Exploration at the Keck Observatory\cite{mosfire} (Keck-MOSFIRE; 10 Jan. 2022 UT). 

For the MMT-Blue observation, we set up a $2''$-wide slit with an effective length of $150''$ toward the image G2 with three different position angles of the slit, and obtained a total integration time of 4.22\,hr. The spectra were obtained using the 300\,line\,mm$^{-1}$ grating, providing a wavelength coverage of 3800--9100\,\AA\ with a spectral resolution (full width at half-maximum intensity, FWHM) of 6.47\,\AA. 
We centred two $1''$ slits at the galaxy images G2 and G3 for our LBT-LUCI observation, and took a total exposure of 1.2\,hr on our targets. We obtained $H$- and $K$-band spectra of G2 and G3 from the LBT-LUCI observation, covering a wavelength range from 15,500\,\AA\ to 25,000\,\AA. 
For our Keck-MOSFIRE observation, we set up a $0.7''$ slit going through the galaxy images G1 and G3, and obtained spectra from 14,500\,\AA\ to 18,000\,\AA\ using an exposure of 0.86\,hr in the $H$ band. 
The MMT-Blue and LBT-LUCI data were reduced using {\tt PypeIt} software\cite{pypeit:joss_arXiv,pypeit:zenodo} and the Keck-MOSFIRE data were reduced using the MOSFIRE Data Reduction Pipeline\cite{mosfire_drp}.

For the MMT-Blue spectroscopy, the spectrum of our target is expected to have two components: one from the bright foreground galaxy and one from the faint G2 image, where most of the continuum is contributed by the brighter component. 
We fit the continuum with {\tt FireFly} software\cite{firefly} using stellar-population models\cite{m11_stellarmodel}.
We find no $>3\sigma$ peaks detected from the residual spectrum.
At $z=3$, we expect the Lyman-$\alpha$ emission line to have an observer-frame wavelength of $4863$\,\AA, and we obtain a $3\sigma$ upper limit on its flux of
$2.2\times10^{-17}\,\mathrm{erg}\,\mathrm{s}^{-1}\,\mathrm{cm}^{-2}$ within the 6.47\,\AA\ spectral bin, corresponding to an upper-limit equivalent width of $\sim 200$\,\AA\ in the observer's frame.

The near-infrared spectra were taken from observations using two nod positions, and the difference spectroscopic imaging from the two positions was used to measure the spectra of our targets.
There is no single peak with $>2\sigma$ significance in the $H+K$ and $H$ bands. A pair of low-significance peaks can be found at $\sim14,840$\,\AA\ from the coadded Keck-MOSFIRE spectrum, which may correspond to the [O~II] doublet at 3726.0\,\AA\ and 3728.8\,\AA\ in the rest frame at $z=2.983$.
To calculate a $3\sigma$ upper limit for the equivalent width, we estimate the continuum from {\it HST} photometry. Our analysis yields a 3$\sigma$ upper bound of $\sim 350$\,\AA\ for the [O~II] doublet, and we compare this to the distribution measured by the MOSDEF survey\cite{mosdef}. The host galaxy of the multiply imaged SN we have discovered has a stellar mass of $\sim 2 \times 10^8$\,$M_\odot$, which is less massive than a majority of the MOSDEF galaxies with detected [O~II] emission lines. Therefore, an [O~II] detection would not be expected, given the MOSDEF sample.

\subsection{Modelling the gravitational lens:}

After the discovery of the transient, we first examined the predictions of the existing HFF lens models\cite{lotz2017}. Since the multiple-image system is only several arcseconds in size, we considered three high-resolution published models with $\leq 0.06''\,\mathrm{pixel}^{-1}$: Keeton V4\cite{mccullykeetonwong14,ammonswongzabludoff14,keeton10}, GLAFIC V4\cite{kawamataishigakishimasaku18,kawamataoguriishigaki16,oguri10}, and Sharon V4\cite{johnsonsharonbayliss14,jullokneiblimousin07}. Among these three models, the Keeton V4 model provides the best reconstruction of the multiple images given a source redshift of $z=3$, as shown in Extended Data Fig.\,3a. However, the mass of the early-type galaxy is not included as a free parameter for these models, since the multiply imaged host galaxy was not yet known. Consequently, we developed our own lens model using {\tt GLAFIC}.

To explore geometric constraints on the SN's redshift independent of our photometric redshift using these ``blind'' models, we next allow the redshift of the SN to be a free parameter. In Extended Data Fig. 3b, we plot the root-mean-square (RMS) angular separation of the predicted images from the observed positions as a function of the source redshift. For all three published models, we are able to rule out low redshifts at $z \approx 1$, where the RMS angular separation of the predicted images is much larger than that at higher redshifts.

In Extended Data Fig. 3b, we can see that, for redshifts exceeding 3, the models still yield relatively small RMS residuals. This is because Einstein radii of the lens, for the published models, are slightly smaller than the true value obtained by allowing the galaxy's mass to be a free parameter. The Einstein radius depends on the ratio of angular diameter distances $D_{ls}/D_{os}$, where $D_{ls}$ and $D_{os}$ are angular diameter distances from the lens to the source and from the observer to the source, respectively. However, this ratio changes slowly for source redshifts beyond $z\approx3$. 
Consequently, the models' underestimated mass for the early-type lens cannot be compensated for by increasing the source's redshift.

To employ newly discovered image positions as constraints and allow the parameters of the early-type lens to vary freely, we modelled the gravitational lens in the Abell~370 cluster using the {\tt GLAFIC} modelling package\cite{oguri10,oguri21} and MCMC analysis.
We explicitly included the observed image positions as part of the fit. We adopted a prior on each image's position within a standard deviation of $0.1''$ as the standard deviation of the prior probability.
We use a truncated SIE model where the velocity dispersion, truncation radius, ellipticity, and position angle are all free parameters independent from those of the other cluster members to model the early-type lens. 
We use the joint probability density we obtained from the photometric-redshift measurement as shown in Extended Data Figs.\,2g and 2h to apply a Gaussian prior on the SN host galaxy's redshift at 2.95 with $\sigma = 0.05$.
As shown in Extended Data Fig.\,3c, our GLAFIC model can reproduce the multiple SN images with RMS residuals of $< 0.02''$.

The original GLAFIC V4 model predicts that the early-type galaxy has a velocity dispersion of the lens of 152\,km\,s$^{-1}$, while our revised model, which is constrained by the image positions and photometric redshift, yields a velocity dispersion of the lens of 172\,km\,s$^{-1}$. The difference between these values is within the scatter of the Faber-Jackson relation\cite{faberjackson}. If we fix the SN redshift to $z=1$, the velocity dispersion of the lens becomes much larger, $\sim268$\,km\,s$^{-1}$, and therefore its mass-to-light ratio significantly deviates from those of the other cluster member galaxies. This analysis provides independent evidence from gravitational lensing alone that excludes low redshifts at $z \approx 1$.

The resulting time delays ($\Delta t$) and magnifications ($\mu$) of the multiple images are listed in Extended Data Table\,1b. We note that the uncertainties associated with $\Delta t$ and $\mu$ listed in Extended Data Table\,1b are the standard deviations of each individual parameter.
These parameters are highly correlated, as shown by the distribution of $\Delta t$ and $\mu$ values from the MCMC samples of the lens model plotted in Extended Data Figs.\,3d and 3e. The distribution of $\Delta t$ and $\mu$ given by the model samples is used as the prior density for the following light-curve fitting.
The magnification contributed by the macro cluster-lens model at the position of the multiple images is $\sim 8$, as given by the sum of the signed magnifications of the quadruple\cite{witt2000}.

\subsection{Light-curve fitting:}

The breakout may take place when the shock reaches the surface of the progenitor star. In that case, the early-time emission from the shock is from the star's envelope. The early light curve of SNe based on this mechanism has been modelled by previous studies\cite{ns10,rw11,sw17}. Alternatively, if an amount of CSM has been ejected prior to the SN explosion and its optical depth is larger than $c/v$, the breakout will occur at a large radius outside the star's surface. In such a case, the early-time SN emission is from the shocked CSM\cite{np14,p15,p21,m21}, in which a low-mass ($\sim0.01$--0.1\,$M_\odot$) and extended ($\sim10^{13}$\,cm) CSM is assumed. Such a scenario could also be used to explain the emission from Fast Blue Optical Transients (FBOTs\cite{droutchornocksoderberg14,hophinneyravi19,marguttimetzgerchornock19})\cite{m21}.

We fit the observed data with shock-cooling models utilising our gravitational lens model. We considered two shock-cooling scenarios assuming shock breakouts in massive envelopes of supergiant stars\cite{sw17} or in light-mass and dense CSM shells around progenitors\cite{p21}.
Detailed descriptions of the models we use can be found in the literature\cite{p21,m21,sn2018fif}. For all models, we assumed the opacity to be 0.34\,cm$^2$\,g$^{-1}$. The bandpass flux is computed using the {\tt pysynphot} package\cite{pysynphot} based on blackbody spectra.

Each of the SN models includes a set of four free parameters,
that comprise of the radius $R$ and mass $M$ of the envelope or the CSM, the shock velocity $v$, and the initial time of the shock breakout $t_0$, which are as follows: $R$, $M$, $v$, $t_0$, $\Delta t_{13}$, $\Delta t_{23}$, $\mu_{S1}$, $\mu_{S2}$, $\mu_{S3}$, and $E(B-V)$, where where $R$ and $M$ are the radius and mass of the envelope or the CSM, $v$ is the shock velocity, $t_0$ is the initial time of the shock breakout, $\Delta t_{13}$ and $\Delta t_{23}$ are (respectively) the relative time delays of S1 and S2 with respect to S3, and $\mu_{S1}$, $\mu_{S2}$, and $\mu_{S3}$ are (respectively) magnifications of S1, S2, and S3. $R$, $M$, and $v$ are on a logarithmic scale in the parameter space.

The reconstruction of the extinction-corrected rest-frame light curve from the observations involves six additional parameters that are constrained by the lens model. The lensing-related parameters are $\Delta t_{13}$ and $\Delta t_{23}$, respectively the relative time delays of S1 and S2 relative to S3, as well as $\mu_{S1}$, $\mu_{S2}$, and $\mu_{S3}$, respectively the magnifications of S1, S2, and S3. Finally, $E(B-V)$ is the colour excess of the SN due to the host-galaxy dust extinction.

We next perform an MCMC analysis using the {\tt emcee}\cite{emcee} package to obtain the posterior probability distribution for the model parameters. 
We adopt the {\tt GLAFIC} posterior probability density distributions as priors on $\Delta t_{13}$, $\Delta t_{23}$, $\mu_{S1}$, $\mu_{S2}$, and $\mu_{S3}$.
For all of the remaining parameters, we implement uniform priors within the parameter ranges listed in Extended Data Table\,1c. The parameter values and their confidence intervals for all models are listed in Extended Data Table\,2.
In our fitting, epochs are defined relative to the rest-frame time when the F814W imaging of S3 was acquired.

The CSM-planar model has additional parameters, including the shock-velocity scale. We chose the same values as used in previous modeling efforts\cite{m21}: $x_0=R_0/7$, $v_0=v_e$, and $\beta=0.5$.
We considered three options of the host-galaxy dust extinction with $R_V=2.74$, $3.1$, or $4$ for the cases of the Small Magellanic Cloud bar, average Milky Way diffuse, and extragalactic starburst, respectively. We assumed that the colour excess from the host-galaxy extinction $E(B-V)$ is a free parameter. Corner plots in Extended Data Fig.\,4 and Fig.\,5 show the distribution of free parameters from the MCMC samples for $R_V=3.1$ for the two models that can reproduce our observations. Light curves from the best-fit models are shown Extended Data Fig.\,6 for $R_V=3.1$. MCMC samples, corner plots, and best-fit light curves for all models with all our choices of $R_V$ values are available in an online repository\cite{supp_repo}.

The first set of photometric measurements of the SN in its rest frame are those made from the image S1, which is closest to the early-type lens. This image has a greater background contribution from the galaxy's light than the other two images, especially in the two near-IR filters. There is no statistically significant detection of the S1 image from the F160W filter, where we measure a flux of $25.19\pm16.45$\,nanoJansky. To explore the effect of the early measurement, we repeated light-curve fitting using the RSG and CSM-homologous (with CSM mass $10^{-3}\,M_\odot < M < 1\,M_\odot$) models after doubling the uncertainty associated with the F160W flux of S1. The constraints on the progenitor radii that result from these fits are $526^{+162}_{-118}\,R_\odot$ and $469^{+158}_{-118}\,R_\odot$, respectively. Even if the uncertainty of our photometry of the most contaminated image is underestimated by a factor of 2, our constraints on the progenitor radius will not be significantly altered.

\subsection{Bayes factors:}

To compare the results from fitting each light-curve model, we evaluated the Bayes factor for each model against a constant-flux model which describes a multiply imaged blackbody with an unvaried temperature that is differently magnified by the galaxy and cluster lenses. For a dataset $D$, the Bayes factor $B_{10}$ of a hypothesis $\mathcal{H}_1$ against the null hypothesis $\mathcal{H}_0$ is given by
\begin{equation}
B_{10}=\frac{\mathcal{L}_B(D|\mathcal{H}_1)}{\mathcal{L}_B(D|\mathcal{H}_0)},
\end{equation}
in which the marginal likelihood $\mathcal{L}_B$ is given by
\begin{equation}
\mathcal{L}_B(D|\mathcal{H})=\int\mathrm{d}xP(D|x,\mathcal{H})\pi(x|\mathcal{H}),
\end{equation}
where $x$ is the set of model parameters, and $P$ and $\pi$ are the likelihood functions and prior densities (respectively).
Prior densities of time-delay and magnification parameters are from the parameter distribution of the MCMC samples given by our lens modelling. We used the same prior densities for the other parameters as for the parameter estimation, whose upper and lower bounds are listed in Extended Data Table\,1c. The integration in the high-dimensional parameter space is evaluated using the Monte Carlo (MC) integration method. In particular, we generated a number of MC samples based on the prior densities. The marginal likelihood $\mathcal{L}_B$ can be approximated by averaging the calculated likelihood over all the MC samples, which is given by
\begin{equation}
\mathcal{L}_B\approx \frac{1}{n}\sum_i^n P(D|x_i,\mathcal{H}), x_i\sim\pi(x_i|\mathcal{H}).
\end{equation}
In this work, we evaluated the marginal likelihood using a numerical integration based on $10^8$ MC samples for each model. We then calculated the normal-logarithm Bayes factor for each shock-cooling model against the constant-flux model. The results are listed in Extended Data Table\,1c. Since the Bayes factor is the ratio of the marginal likelihoods of two competing statistical models, the difference between two logarithmic Bayes factors corresponds to the models' logarithmic Bayes factor. 

\subsection{Microlensing optical depth:}

The colours of multiple images of a single source can become different if the caustic of a microlens, such as a star or compact object in the foreground lens, intersects the expanding SN photosphere. 
If the scale of the source-plane caustic from microlenses is comparable to the size of the source's photosphere, photospheric regions with different temperatures could be magnified differently, leading to a colour change in the microlensed images. The probability of microlensing can be estimated from the optical depth for microlensing. We measured the photometry of the intracluster light (ICL) in the vicinity of the three images, and then fit it using the {\tt FAST++} software\cite{fast,fastpp} to obtain the stellar mass. We obtained a surface stellar mass density of $3.55\,M_\odot\,\mathrm{pc}^{-2}$, $3.47\,M_\odot\,\mathrm{pc}^{-2}$, and $2.09\,M_\odot\,\mathrm{pc}^{-2}$ near G1, G2, and G3, respectively. The result is consistent with the surface mass density\cite{morishitaabramsontreu17} at a few hundred kpc away from the centre of the Abell~370 cluster.
The optical depth of microlensing\cite{diegokaiserbroadhurst18} can be given by
\begin{equation}
\begin{aligned}
\tau=\int_0^{D_L}\Omega_En(D_L)dD_L,
\end{aligned}
\label{ml_optiocal_depth}
\end{equation}
where $\Omega_E$ is the solid angle covered by the Einstein ring from microlenses, and $n(D_L)$ is the number density of the microlenses at an angular-diameter distance $D_L$. To simplify the calculation, we adopt a thin-lens assumption that all mass along the line of sight is concentrated in the cluster and assume all microlenses have the same mass that generate an Einstein ring with radius $\theta_E$. Eq. \ref{ml_optiocal_depth} can then be simplified as
\begin{equation}
\begin{aligned}
\tau\approx\mu\pi \theta_E^2n_L,
\end{aligned}
\label{ml_optiocal_depth_simple}
\end{equation}
where $\mu$ is the magnification of the macro lens from the cluster and $n_L$ is the number density of the microlenses in the cluster lens. Plugging the surface star density (assuming solar-mass microlenses) and the magnification we obtained earlier, we have the microlensing optical depth of 0.010, 0.013, and 0.006 for S1, S2, and S3, respectively. The optical depth is very small, indicating that the probability that at least two images are significantly microlensed is negligible. 

\subsection{Comparison with UV light curves of low-redshift SNe:}

In Extended Data Fig.\,7a, we compare the evolution of the effective blackbody temperature of the newly discovered SN with those of an RSG model, SN~2018fif\cite{sn2018fif} (Type IIP, with a $\sim700\,R_\odot$ RSG progenitor\cite{sn2018fif}), SN~2013ej\cite{valenti2014} (Type IIP, with a 400--600\,$R_\odot$ RSG progenitor\cite{valenti2014}, or a $2100\,R_\odot$ CSM shell\cite{morozova2017}), SN~2017eaw\cite{szalai2019,rui2019} (Type IIP, with a $\sim600\,R_\odot$ RSG progenitor\cite{rui2019}), SN~1987A\cite{sn1987a}, and SN~2016gkg\cite{sn2016gkg,tartaglia2017,arcavi2017,p21} (Type IIb, with the early-time emission driven by a $\sim0.03\,M_\odot$ CSM shell\cite{p21}). We can see that the early-time temperature of this event is consistent with the examples of Type IIP SNe.

Modelling of the light curves of SNe~IIP finds evidence that a significant fraction of the progenitors of CCSNe have an extended, massive CSM\cite{morozova2017}. 
We note that the early-time light curve of an SN~IIP is only sensitive to CSM that has an optical depth exceeding $c/v$. 
We next compare our event's light curve with those of two recently discovered SNe~IIP/L, SN~2018fif\cite{sn2018fif} and SN~2021yja\cite{sn2021yja}, whose early light curves allow only a small CSM mass ($\lesssim 0.001\,M_{\odot}$) around their progenitors\cite{sn2018fif,sn2021yja}. In the rest frame of our SN at $z = 3$, the central wavelengths of the ACS-WFC F814W, WFC3-IR F110W, and WFC3-IR F160W filters correspond approximately to the {\it Swift}-UVOT's $B$, $UVW1$, and $UVW2$ bands, respectively. As shown in Extended Data Fig.\,7b, the early light curve of our multiply imaged SN is consistent with the {\it Swift}-UVOT observations of SN~2018fif and SN~2021yja.

Analytical models show that very early UV observations at wavelengths shorter than 3000\,\AA\ are needed to constrain the SN's temperature, which is important to accurately infer the radius where the shock breaks out\cite{sw17}. As shown in Extended Data Figs.\,7c and 7d, we construct SN~IIP light curves using the analytical models with CSM at $R<1000\,R_{\odot}$ and $R>1000\,R_{\odot}$. Recent analysis finds evidence that a larger fraction RSG progenitors of SNe~IIP may have CSM shells, but CSM with large masses ($M\,\gtrsim\,0.05\,M_{\odot}$) are only found at $R \gtrsim 1000\,R_{\odot}$ \cite{morozova2017}. Our analytical light curves plotted in Extended Data Fig.\,7 show that a photosphere or CSM (with optical depth $>c/v$) smaller than $\sim 1000\,R_\odot$ causes the UV luminosity to fall more rapidly following shock breakout, with subsequent rebrightening, while the UV light curves for objects with more extended CSM exhibit much more modest rebrightening.

As Extended Data Fig. 7 demonstrates, detection of rebrightening requires UV observations within the first day in the rest frame. While our dataset includes these very early UV observations, they are absent from almost all low-redshift SNe~IIP.  We show the light curves and CSM-homologous models of SN~2013fs\cite{sn2013fs} and ASASSN-14gm\cite{valenti2016}, which have been inferred to have CSM shells around their progenitors at $R>1000\,R_{\odot}$\cite{morozova2017,morozova2018}.
Overplotted CSM-homologous models for the CSM parameters for these events\cite{morozova2017,morozova2018} show that UV observations within the first day would be brighter than the measured luminosities for our SN.
We note that, in those analyses of CCSNe with CSM\cite{morozova2017,morozova2018}, the UV data were not used to constrain the CSM properties.

In Extended Data Figs.\,7c and 7d, we plot the observations of our event and the light curves of our best-fitting models (RSG and CSM-homologous). Comparison with the predicted light curves for the CSM-homologous model for the CSM radius and mass\cite{morozova2017,morozova2018}, for SN~2013fs and ASASSN-14gm, demonstrates the connection between the radius of optically dense material and the early UV light curve.
When we rerun light-curve fitting without the first rest-frame UV observation, our constraints become consistent with a much larger set of progenitor radii, $952_{-398}^{+529}\,R_\odot$ from the CSM-homologous model, and favour a value approximately twice as large ($\sim 480\,R_\odot$) as we obtained from the CSM-homologous model with the first rest-frame UV observation. 
In summary, the radius of the optically thick material for our progenitor system is strongly constrained by the earliest UV data points.

Extended Data Figs.\,7e and 7f show that the early light curve of our event is not well matched by those of SNe Types Ic, IIb, and Ia, and also of FBOTs, with available early-time {\it Swift} photometry\cite{sn2020bvc,tartaglia2017,sn2018gv,ho2020}.

The earliest emission from an SN originates from the surface of the exploding progenitor. The differences between the gravitational potential of the photosphere at the observed epochs do not suffice to produce a detectable colour change. For example, applying Birkhoff's theorem, for a BSG with $R=10\,R_\odot$ and $M=20\,M_\odot$, the gravitational redshift of light from the BSG surface is $4.2\times10^{-6}$ for a distant observer. Even for a compact white-dwarf progenitor with $R=0.01\,R_\odot$ and $M=1\,M_\odot$, the gravitational redshift is only $2.1\times10^{-4}$ for light from the object's surface. We also note that, since the light from the background SN travels into and then out of the potential, the SN’s light should not experience any net gravitational redshift.

\subsection{Core-collapse SN rate from multiply imaged SNe:}

To simulate the {\it HST} detection of multiply imaged SNe behind the six HFF galaxy clusters, we randomly generated CCSNe in comoving volumes behind the cluster lens within a $0.03^\circ\times0.03^\circ$ field of view and simulate strong-lensing effects based on the GLAFIC lens models of the six clusters\cite{kawamataishigakishimasaku18,kawamataoguriishigaki16,oguri10,oguri21}. We used the {\tt sncosmo}\cite{sncosmo} package to synthesise the multi-band light curves for each SN. For each type of CCSN, we adopted the mean and standard deviation of its peak $B$-band absolute magnitude from a previous study\cite{richardson2014}. We estimated the significance of any detection of the SN using the actual {\it HST} exposures of archival imaging of the HFF clusters employing the HST Exposure Time Calculators\cite{hst_etc}. In a redshift range from $z_1$ to $z_2$, the expected number of detectable multiply imaged CCSNe in the SN rest-frame time $t$ is given by
\begin{equation}
    n_\mathrm{mCC} = d_H^3\theta^2\int_{z_1}^{z_2}t(z)\beta(z)R_\mathrm{CC}(z)\left(\int_0^z\frac{dz'}{E(z')}\right)^2\frac{dz}{E(z)},
    \label{nmcc}
\end{equation}
where $\theta$ is the angular size of the searching window, $d_H = c/{\rm H}_0$ is the Hubble distance, $R_\mathrm{CC}$ is the volumetric CCSN rate, and $E(z)=\sqrt{\Omega_m(1+z)^3+\Omega_\Lambda}$.

For an observer-frame survey time $t_\mathrm{obs}$, the rest-frame time follows as $t(z) = t_\mathrm{obs}/(1+z)$. For a threshold of the signal-to-noise ratio (SNR) on a point-source detection, $\beta(z)$ is the fraction of the detectable multiply imaged SNe to the total number of the simulated SNe at redshift $z$.
In this work, we assume the detectable SN image has a point-source SNR larger than $5\sigma$, which is the smallest SNR of the detected images of the known multiply imaged SNe. We also require that the detectable multiply imaged SN has at least one image brighter than the threshold. 

In our simulation, we use an exponential $A_V$ distribution that follows $P(A_V)\propto e^{-\lambda_V A_V}$. The CANDELS+CLASH analysis\cite{strolger15} used $\lambda_V=0.187$, although we are not able to reproduce this prior from the data used to construct it\cite{schmidt1994,hamuy2002}, while another study\cite{taylor2014} gave $\lambda_V=1/\tau_V\approx2$ based on simulated CCSN host galaxies\cite{hatano1998}. For a few hundred CCSN host galaxies in the low-redshift universe\cite{kelly2012}, we obtained the best-fit $\lambda_V=0.98\pm0.03$, as shown in Extended Data Fig.\,8a.
For $\lambda_V=1$ and $R_V=4.05$, the mean and median host-galaxy $E(B-V)$ are 0.25 and 0.17\,mag, respectively, consistent with the mean and median $E(B-V)$ from previous studies\cite{drout2011,prentice2016}. In this study, we used $\lambda_V=1$ and the $R_V=4.05$\cite{calzetti00} as the primary values. We have also repeated our simulation with $R_V=3.1$, as well as $\lambda_V=0.187$ and $\lambda_V=2$ as used in the two previous analyses\cite{strolger15,taylor2014}, for the purpose of comparison.

Extended Data Fig.\,8b shows the differential comoving volume for the strongly lensed sources as a function of redshift. Although the lensing volume remains large at high redshifts, we can see that the effective volume for $>5\sigma$ detection decreases with increasing redshift in the high-redshift regime. The resulting differential number of multiply imaged CCSNe with $>5\sigma$ detection significance by {\it HST} in the last decade is shown in Extended Data Fig.\,8c. At high redshifts, the number of detectable multiply imaged SNe is small, not only because the effective lensing volume for detectable sources decreases with increasing redshift, but also because of the redshift time dilation. For a given survey period in the observer's frame, the corresponding rest-frame time at higher redshifts is much shorter than that at lower redshifts.

We chose four redshift bins: 0.5--1, 1--2, 2--3.5, and 3.5--6. The number of detected multiply imaged CCSNe ($N_\mathrm{mCC}$) within each redshift bin follows a Poisson distribution as $Poi(N;n)=n^N e^{-n}/N!$, where $n=n_\mathrm{mCC}$ and $N=N_\mathrm{mCC}$. As shown in Fig.\,4, we constrain the average $R_\mathrm{CC}$ for each redshift bin to the 68\% confidence level based on the two discovered multiply imaged SNe, where upper limits are for redshift bins with nondetection.

The coefficient $k_\mathrm{CC}$ can be calculated to be $0.0068~M_\odot^{-1}$~\cite{MD14} for a Salpeter IMF\cite{salpeter_imf} and an initial mass range of CCSN progenitors 8--40\,$M_{\odot}$. Adopting the star-formation rate density from an analysis of the cosmic star-formation history (CSFH) inferred from observations of galaxies\cite{MD14},
\begin{equation}
\begin{aligned}
\psi(z) = \frac{A(1+z)^C}{((1+z)/B)^D + 1}\,M_\odot\,\mathrm{yr}^{-1}\,\mathrm{Mpc}^{-3},
\end{aligned}
\label{sfr}
\end{equation}
where $A=0.015$, $B=2.9$, $C=2.7$, and $D=5.6$, we reproduce the previously predicted CCSN rate\cite{MD14}, as shown by the solid black curve in Fig.\,4 (the same as the curve shown in Fig.\,10 of the CSFH analysis\cite{MD14}).

On the other hand, for a given model of star-formation rate density ($\psi$), $k_\mathrm{CC}$ can also be constrained using the total number of detected multiply imaged CCSNe behind the six HFF clusters. Here we evaluate a posterior probability density as given by
\begin{equation}
P(k_\mathrm{CC})=Poi\left(N_\mathrm{mCC};n_\mathrm{mCC}\right)\times\sqrt{1/n_\mathrm{mCC}},
\label{p_kcc}
\end{equation}
where $\sqrt{1/n_\mathrm{mCC}}$ is from the Jeffreys prior\cite{jeffreys_prior} for Poisson distributed variables. The variable $n_\mathrm{mCC}$ is a function of $R_\mathrm{CC}$, and therefore is a function of $k_\mathrm{CC}$ and $\psi$,
which can be derived using Eqs.~\ref{nmcc} and \ref{sfr}. Using Eq.~\ref{sfr} and the parameters from the analysis of the CSFH\cite{MD14}, we find $k_\mathrm{CC}=0.0087_{-0.0046}^{+0.0071}\,M_\odot^{-1}$ and a MLE value of $k_\mathrm{CC}=0.0060\,M_\odot^{-1}$.

SN Refsdal, the first-known example of a strongly lensed SN, and one of the most distant examples of an SN (at $z = 1.49$), was classified as SN~1987A-like\cite{kelly2016}. The classification was surprising, because SN~1987A-like SNe account for only 1--3\% of CCSNe reported by low-redshift surveys\cite{pastorello2012,taddia2016}. The Caltech Core-Collapse Project  reported a somewhat greater percentage of 5\%\cite{taddia2016}.

We used our simulations to estimate the probability that, of the two first discoveries of strongly lensed CCSNe (by a cluster lens), at least one was SN~1987A-like.  The absolute magnitudes of the SN~1987A-like SNe in our simulations have a mean of $-16$ and standard deviation of unity, as an approximation of the distribution of absolute magnitudes of SN~1987A-like SNe\cite{kelly2016}. As this figure shows, SN Refsdal would be a relatively luminous SN~1987A-like SN with an absolute $B$-band magnitude of approximately $-16.5$ given a magnification of 15 for images S1–-S3. Assuming that SN~1987A-like SNe account for 5\% of CCSNe to $z \approx 4$, we find that we should expect 1.7 strongly lensed CCSNe in the existing observations of the HFF cluster fields, but only 0.09 of these should be SN~1987A-like SNe. Consequently, the discovery of SN Refsdal, an SN~1987A-like SN\cite{kelly2016} at $z=1.49$, is not highly probable. However, the $p$ value exceeds $p=0.05$, given the two known examples.

We have rerun our simulation with an extreme assumption that SN~1987A-like SNe account for 30\% of CCSNe in the lensed volume, and find that we should expect 1.76 strongly lensed SNe in the existing observations of the HFF cluster fields. Consequently, inferences about the core-collapse rates are almost entirely unchanged (only $\sim3.5$\%), indicating that
our constraints will not be significantly affected by our assumptions about the rate of SN~1987A-like SNe at high redshift.

As shown in Extended Data Fig.\,8e, our constraints on the volumetric rate of CCSNe obtained using $R_V=3.1$ are not significantly different from the those derived with $R_V=4.05$. However, our estimate of the CCSN rate changes significantly with the parameter $\lambda_V$ of the exponentially distributed $A_V$.
For $\lambda_V=0.187$, the same as used in a previous analysis of high-redshift SN rates\cite{strolger15}, we find $k_\mathrm{CC}=0.0216_{-0.0111}^{+0.0146}\,M_\odot^{-1}$ and the MLE $k_\mathrm{CC}=0.0162\,M_\odot^{-1}$.
After correction for an extraneous factor of $h^2 = 0.7^2$ (see discussion below), our reanalysis of the CANDELS+CLASH analysis\cite{strolger15} finds that their measurements yield $k_\mathrm{CC}=0.0045\pm0.0008\,M_\odot^{-1}$.

The volumetric CCSN rate can be expressed as a function of the star-formation density: $R_\mathrm{CC}(z) = \psi(z)\, k_\mathrm{CC}$\cite{MD14}. The CANDELS+CLASH analysis introduced a factor of $h^2$ where H$_0=100h\,\mathrm{km\,s^{-1}\,Mpc^{-1}}$ on the right-hand side of the equation\cite{strolger15}, which we believe is not needed.  
Instead, the equation for the measured $\left<R_\mathrm{CC}\right>$ and $\left<\psi\right>$ should be
\begin{equation}
\left<R_\mathrm{CC}\right>=k_\mathrm{CC}\times\left<\psi\right>\times\left(\frac{{\rm H}_0^*}{{\rm H}_0}\right)^2,
\label{corrected_eq8}
\end{equation}
where H$_0^*$ is the value of H$_0$ used for the star-formation density. However, both the volumetric SN rate from the CANDELS+CLASH analysis\cite{strolger15} and the star-formation density from the analysis of the CSFH\cite{MD14} adopted H$_0=70\,\mathrm{km\,s^{-1}\,Mpc^{-1}}$.  Consequently, the factor ${{\rm H}_0^*}/{{\rm H}_0}$ should be equal to 1.
Following Eq.\,\ref{corrected_eq8}, we find that the $k_\mathrm{CC}$ value reported in the CANDELS+CLASH analysis\cite{strolger15} should be reduced by a factor of $0.7^2$.  

We next compute the probability of detecting the SN we have found at $z \approx 3$, given the constraints on the CCSN rate from the CANDELS+CLASH analysis\cite{strolger15}. The authors used the functional form for the cosmic star-formation history\cite{MD14}, as shown in Eq.\ref{sfr}, but fit for new values of the coefficients: $A=0.015$, $B=1.5$, $C=5.0$, and $D=6.1$. We simulate multiply imaged SNe based on the cosmic star-formation history, the value of $k_\mathrm{CC}$, and the extinction parameter $\lambda_V=0.187$\cite{strolger15}. Our simulations show that we expect to detect $n_\mathrm{mCC}=0.25$ multiply imaged SNe in the existing observations of the HFF cluster fields based on the parameters from the CANDELS+CLASH analysis\cite{strolger15}. We find a small probability of detecting two or more multiply imaged SNe, $p(N_\mathrm{mCC}\ge2)=0.026$. Moreover, the star-formation history from the CANDELS+CLASH analysis\cite{strolger15} declines rapidly in the range $z \approx 1$--3, and the probability of finding at least one multiply imaged SNe at $z \ge 3$ is very small ($p \approx 0.01$). The detection of our event is extremely unlikely given the constraints from the CANDELS+CLASH analysis\cite{strolger15}.
In Extended Data Fig.\,8e, we compare our constraints on the volumetric CCSN rate with those from the CANDELS+CLASH analysis\cite{strolger15}.

For $\lambda_V=2$\cite{taylor2014}, the estimated CCSN rate is smaller than what we obtained from $\lambda_V=1$ based on the  distribution of $A_V$ measured for the host galaxies of nearby SNe \cite{kelly2012}, but the 68\% confidence intervals for these two $\lambda_V$ choices, as shown by the green-shaded region and the blue-shaded region in Extended Data Fig.\,8e, highly overlap, indicating little statistical difference. 
For $\lambda_V=2$, we obtained $k_\mathrm{CC}=0.0072_{-0.0038}^{+0.0059}\,M_\odot^{-1}$ and the MLE $k_\mathrm{CC}=0.0049\,M_\odot^{-1}$ from observed multiply imaged SNe,
based on the cosmic star-formation history from analysis of the CSFH\cite{MD14}.

To date, we have only reanalysed archival imaging of the six HFF clusters, which led to the discovery of the new strongly lensed SN not identified by previous searches.
In the earlier CLASH program, imaging of 25 galaxy-cluster fields (including four of the six HFFs) was acquired with a total of 524 orbits. No multiply imaged SN has been publicly reported from the CLASH data in the previous studies. If we assume that none was missed and an SN can be detected within a 0.5\,yr time period bracketing each observation, our volumetric CCSN rate would decrease by a factor of $\sim2$, yielding $k_\mathrm{CC} \approx 0.008~M_\odot^{-1}$ for the host-galaxy extinction prior from the CANDELS+CLASH analysis\cite{strolger15}. Nonetheless, our CCSN rate and the $k_\mathrm{CC}$ value are still higher than the constraints obtained from blank-field {\it HST} surveys\cite{strolger15}. 
If we assume that no multiply imaged SNe from the CLASH survey were missed, then the tension between our measurements and those of the CANDELS+CLASH analysis\cite{strolger15} (after correction) diminishes to $p=0.088$, but the probability of finding at least one multiply imaged SN at $z \ge 3$ is still small ($p \approx 0.02$).

\end{methods}


\begin{thebibliography}{1}
	
	\bibitem{waxman17}
	Waxman, E. \& Katz, B. {Shock breakout theory.} {\it Handbook of Supernovae\/} p. 967–1015 (2017).
	
	\bibitem{garnavich2016}
	Garnavich, P. M., {\it et~al.\/}, {Shock breakout and early light curves of Type II-P supernovae observed with Kepler.} {\it Astrophys. J. Letters\/} {\bf 820}, 23 (2016).
	
	\bibitem{sn2011dh-1}
	Arcavi, I., {\it et~al.\/}, {SN 2011dh: discovery of a Type IIb supernova from a compact progenitor in the nearby galaxy M51.} {\it Astrophys. J. Letters\/} {\bf 742}, L18 (2011).
	
	\bibitem{sn2011fu}
	Morales-Garoffolo, A., {\it et~al.\/}, {SN 2011fu: a type IIb supernova with a luminous double-peaked light curve.} {\it Mon. Not. R. Astron. Soc.\/} {\bf 454}, 95-114 (2015).
	
	\bibitem{ptf12gzk}
	Ben-Ami, S., {\it et~al.\/}, {Discovery and early multi-wavelength measurements of the energetic Type Ic supernova PTF12gzk: a massive-star explosion in a dwarf host galaxy.} {\it Astrophys. J. Letters\/} {\bf 760}, L33 (2012).
	
	\bibitem{valenti2014}
	Valenti, S., {\it et~al.\/}, {The first month of evolution of the slow-rising Type IIP SN 2013ej in M74.} {\it Mon. Not. R. Astron. Soc.: Letters\/} {\bf 438}, L101-L105 (2014).
	
	\bibitem{sn2016gkg}
	Bersten, M. C., {\it et~al.\/}, {A surge of light at the birth of a supernova.} {\it Nature\/} {\bf 554}, 497-499 (2018).
	
	\bibitem{tartaglia2017}
	Tartaglia1, L., {\it et~al.\/}, {The progenitor and early evolution of the type IIb SN 2016gkg.} {\it Astrophys. J. Letters\/} {\bf 836}, L12 (2017).
	
	\bibitem{arcavi2017}
	Arcavi, I., {\it et~al.\/}, {Constraints on the progenitor of SN 2016gkg from its shock-cooling light curve.} {\it Astrophys. J. Letters\/} {\bf 837}, L2 (2017).
	
	\bibitem{p21}
	Piro, A. L., Haynie, A., \& Yao, Y., {Shock cooling emission from extended material revisited.} {\it Astrophys. J. Letters\/} {\bf 909}, 209 (2021).
	
	\bibitem{szalai2019}
	Szalai, T., {\it et~al.\/}, {The Type II-P supernova 2017eaw: from explosion to the nebular phase.} {\it Astrophys. J.\/} {\bf 876}, 19 (2019).
	
	\bibitem{rui2019}
	Rui, L., {\it et~al.\/}, {Probing the final-stage progenitor evolution for Type IIP Supernova 2017eaw in NGC 6946.} {\it Mon. Not. R. Astron. Soc.\/} {\bf 485}, 1990-2000 (2019).
	
	\bibitem{xiang2019}
	Xiang, D., {\it et~al.\/}, {Observations of SN 2017ein reveal shock breakout emission and a massive progenitor star for a Type Ic Supernova.} {\it Astrophys. J.\/} {\bf 871}, 176 (2019).
	
	\bibitem{sn2018fif}
	Soumagnac, M. T., {\it et~al.\/}, {SN 2018 fif: the explosion of a large red supergiant discovered in its infancy by the Zwicky Transient Facility.} {\it Astrophys. J. Letters\/} {\bf 902}, 6 (2020).
	
	\bibitem{lotz2017}
	Lotz, J. M., {\it et~al.\/}, {The frontier fields: survey design and initial results.}, {\it Astrophys. J.\/} {\bf 837}, 97 (2017).
	
	\bibitem{strublerood1991}
	Struble, M. F. \& Rood, H. J., {A compilation of redshifts and velocity dispersions for Abell clusters (Epoch 1991.2).}, {\it Astrophys. J. Supplement Series\/} {\bf 77}, 363 (1991).
	
	\bibitem{bpz1}
	Ben{\'\i}tez, N., {Bayesian photometric redshift estimation.} {\it Astrophys. J.\/} {\bf 536}, 571 (2000).
	
	\bibitem{eazy}
	Brammer, G. B., van Dokkum, P. G., \& Coppi, P., {EAZY: A Fast, public photometric redshift code.} {\it Astrophys. J.\/} {\bf 686}, 1503 (2008).	
	
	\bibitem{kawamataishigakishimasaku18}
	Kawamata, R., {\it et~al.\/}, {Size-luminosity relations and UV luminosity functions at z = 6-9 simultaneously derived from the complete Hubble Frontier Fields data.} {\it Astrophys. J.\/} {\bf 855}, 4	(2018).
	
	\bibitem{kawamataoguriishigaki16}
	Kawamata, R., Oguri, M., Ishigaki, M., Shimasaku, K., \& Ouchi, M., {Precise strong lensing mass modeling of four Hubble Frontier Field clusters and a sample of magnified high-redshift galaxies.} {\it Astrophys. J.\/} {\bf 819}, 114 (2016).
	
	\bibitem{oguri10}
	Oguri, M., {The mass distribution of SDSS J1004+4112 revisited.} {\it Publications of the Astronomical Society of Japan\/} {\bf 62}, 1017 (2010).
	
	\bibitem{oguri21}
	Oguri, M., {Fast Calculation of Gravitational Lensing Properties of Elliptical Navarro-Frenk-White and Hernquist Density Profiles.} {\it Publications of the Astronomical Society of the Pacific\/} {\bf 133}, 074504 (2021).
	
	\bibitem{ns10}
	Nakar, E. \& Sari, R., {Early supernovae light curves following the shock breakout.} {\it Astrophys. J.\/} {\bf 725}, 904 (2010).
	
	\bibitem{rw11}
	Rabinak, I. \& Waxman, E., {The early UV/optical emission from core-collapse supernovae.} {\it Astrophys. J.\/} {\bf 728}, 63 (2011).
	
	\bibitem{np14}
	Nakar, E. \& Piro, A. L., {Supernovae with two peaks in the optical light curve and the signature of progenitors with low-mass extended envelopes.} {\it Astrophys. J.\/} {\bf 788}, 193 (2014).
	
	\bibitem{p15}
	Piro, A. L., {Using double-peaked supernova light curves to study extended material.} {\it Astrophys. J. Letters\/} {\bf 808}, L51 (2015).
	
	\bibitem{sw17}
	Sapir, N. \& Waxman, E., {UV/optical emission from the expanding envelopes of Type II supernovae.} {\it Astrophys. J.\/} {\bf 838}, 130 (2017).
	
	\bibitem{morozova2017}
	Morozova, V., Piro, A. L., \& Valenti, S., {Unifying Type II Supernova light curves with dense circumstellar material.} {\it Astrophys. J.\/} {\bf 838}, 28 (2017).
	
	\bibitem{morozova2018}
	Morozova, V., Piro, A. L., \& Valenti, S., {Measuring the progenitor masses and dense circumstellar material of Type II supernovae.} {\it Astrophys. J.\/} {\bf 858}, 15 (2018).
	
	\bibitem{m21}
	Margalit, B., {Analytic light curves of dense CSM shock breakout and cooling.} {\it Astrophys. J.\/} {\bf 933}, 238 (2022).
	
	\bibitem{kass1995}
	Kass, R. E. \& Raftery, A. E., {Bayes factors.}	{\it Journal of the American Statistical Association\/} {\bf 90}:430, 773-795 (1995).
	
	\bibitem{li2011}
	Li, W, {\it et~al.\/}, {Nearby supernova rates from the Lick Observatory Supernova Search – II. The observed luminosity functions and fractions of supernovae in a complete sample.} {\it Mon. Not. R. Astron. Soc.\/} {\bf 412}, 1441-1472 (2011).
	
	\bibitem{kellyrodneytreu15}
	Kelly, P. L., {\it et~al.\/}, {Multiple images of a highly magnified supernova formed by an early-type cluster galaxy lens.} {\it Science\/} {\bf 347}, 1123 (2015).
	
	\bibitem{goobar2017}
	Goobar, A., {\it et~al.\/}, {iPTF16geu: A multiply imaged, gravitationally lensed type Ia supernova.} {\it Science\/} {\bf 356}, 291-295 (2017).
	
	\bibitem{rodney2021}
	Rodney, S. A., {\it et~al.\/}, {A gravitationally lensed supernova with an observable two-decade time delay.} {\it Nature Astronomy}, (2021).
	
	\bibitem{oguri2019}
	Oguri, M., {Strong gravitational lensing of explosive transients.} {\it Reports on Progress in Physics\/} {\bf 82}, 126901 (2019)
	
	\bibitem{foxleymarrable2020}
	Foxley-Marrable, M., {\it et~al.\/}, {Observing the earliest moments of supernovae using strong gravitational lenses.} {\it Mon. Not. R. Astron. Soc.\/} {\bf 495}, 4622-4637 (2020).
	
	\bibitem{candels}
	Grogin, N. A., {\it et~al.\/}, {CANDELS: The Cosmic Assembly Near-infrared Deep Extragalactic Legacy Survey} {\it Astrophys. J. Supplement\/} {\bf 197}, 35 (2011).
	
	\bibitem{clash}
	Postman, M., {\it et~al.\/}, {The cluster lensing and supernova survey with Hubble: an overview.} {\it Astrophys. J. Supplement\/} {\bf 199}, 25 (2012).
	
	\bibitem{strolger15}
	Strolger, L. G., {\it et~al.\/}, {The rate of core collapse supernovae to redshift 2.5 from the CANDELS and CLASH supernova surveys.} {\it Astrophys. J.\/} {\bf 813}, 93 (2015).
	
	\bibitem{schmidt1994}
	Schmidt, B. P., {\it et~al.\/}, {The distance of five Type II supernovae using the expanding photosphere method and the value of H0.} {\it Astrophys. J.\/} {\bf 432}, 42 (1994).
	
	\bibitem{hamuy2002}
	Hamuy, M. \& Pinto, P. A., {Type II supernovae as standardized candles.} {\it Astrophys. J.\/} {\bf 566}, L63 (2002).
	
	\bibitem{kelly2012}
	Kelly, P. L. \& Kirshner, R. P. {Core-collapse supernovae and host galaxy stellar populations.} {\it Astrophys. J.\/} {\bf 759}, 107 (2012).
	
	\bibitem{drout2011}
	Drout, M. R., {\it et~al.\/}, {The first systematic study of Type Ibc supernova multi-band light curves.} {\it The Astronomical Journal\/} {\bf 741}, 97 (2011).
	
	\bibitem{prentice2016}
	Prentice, S. J., {\it et~al.\/}, {The bolometric light curves and physical parameters of stripped-envelope supernovae.} {\it Mon. Not. R. Astron. Soc.\/} {\bf 458}, 2973-3002 (2016).
	
	\bibitem{MD14}
	Madau, P. \& Dickinson, M., {Cosmic star-formation history.} {\it Annual Review of Astronomy and Astrophysics\/} {\bf 52}, 415-486 (2014).
	
	\bibitem{salpeter_imf}
	Salpeter, E. D., {The Luminosity Function and Stellar Evolution.} {\it Astrophys. J.\/} {\bf 121}, 161 (1955).

    \setcounter{mybibstartvalue}{\value{enumiv}}
    
\end{thebibliography}

\begin{thebibliography}{1}
	
	\setcounter{enumiv}{\value{mybibstartvalue}}
	
    \bibitem{fruchter10}
	Fruchter, A. S., Hack, W., Dencheva, M., Droettboom, M., \& Greenfield, P.,
	{BetaDrizzle: A redesign of the MultiDrizzle package.} {\it 2010 Space Telescope Science Institute Calibration Workshop, p. 382-387\/} (2010).
	
	\bibitem{jonesscolnicrodney15}
	Jones, D. O., Scolnic, D. M., \& Rodney, S. A., {PythonPhot: Simple DAOPHOT-type photometry in Python.}, {\it Astrophys. Source Code Lib.} (2015).
	
	\bibitem{stetson87}
	Stetson, P. B., {DAOPHOT - A computer program for crowded-field stellar photometry.} {\it Publications of the Astronomical Society of the Pacific\/} {\bf 99}, 191 (1987).
	
	\bibitem{galfit}
	Peng, C. Y., Ho, L. C., Impey, C. D., \& Rix, H. W., {Detailed structural decomposition of galaxy images.} {\it The Astronomical Journal\/} {\bf 124}, 266 (2002).
	
	\bibitem{dong2016}
	Dong, S., {\it et~al.\/}, {ASASSN-15lh: A highly super-luminous supernova.} {\it Science\/} {\bf 133}, {\bf 351, 6270}, 257-260 (2016).
	
	\bibitem{bradacmaruhuang19}
	Brada{\v{c}}, M., {\it et~al.\/}, {Hubble Frontier Field photometric catalogues of Abell 370 and RXC J2248.7-4431: multiwavelength photometry, photometric redshifts, and stellar properties.} {\it Mon. Not. R. Astron. Soc.\/} {\bf 489}, 99 (2019).
	
	\bibitem{m11_stellarmodel}
	Maraston, C. \& Str{\"o}mb{\"a}ck, G., {Stellar population models at high spectral resolution.} {\it Mon. Not. R. Astron. Soc.\/} {\bf 418}, 2785 (2011).
	
	\bibitem{schulze2021}
	Schulze, S., {\it et~al.\/}, {The Palomar Transient Factory core-collapse supernova host-galaxy sample. I. Host-galaxy distribution functions and environment dependence of core-collapse supernovae.} {\it The Astronomical Journal Supplement Series\/} {\bf 255}, 29 (2021).
	
	\bibitem{mmtblue}
	Schmidt, G. D., Weymann, R. J., \& Foltz, C. B., {A moderate-resolution, high-throughput CCD channel for the MMT spectrograph.} {\it Publications of the Astronomical Society of the Pacific\/} {\bf 101}, 713 (1989).
	
	\bibitem{luci}
	Rothberg, B., {\it et~al.\/}, {Current status of the facility instrumentation suite at the Large Binocular Telescope Observatory.} {Proceedings Volume 10702}, {Ground-based and Airborne Instrumentation for Astronomy VII}, 1070205 (2018).
	
	\bibitem{mosfire}
	McLean, I. S., {\it et~al.\/}, {MOSFIRE, the multi-object spectrometer for infrared exploration at the Keck Observatory.} {Proc. SPIE 8446}, {Ground-based and Airborne Instrumentation for Astronomy IV}, 84460J (2012).
	
	\bibitem{pypeit:joss_arXiv}
	Prochaska, J. X., {\it et~al.\/}, {PypeIt: the Python spectroscopic data reduction pipeline.} Preprint at arXiv:2005.06505 (2020).
	
	\bibitem{pypeit:zenodo}
	Prochaska, J. X., {\it et~al.\/}, {pypeit/PypeIt: Release 1.0.0.} Zenodo 
	
	https://doi.org/10.5281/zenodo.3743493 (2020).
	
	\bibitem{mosfire_drp}
	Konidaris, N. \& Steidel, C. {MOSFIRE DRP.} 
	
	https://keck-datareductionpipelines.github.io/MosfireDRP/\#mosfire-drp (2018).
	
	\bibitem{firefly}
	Wilkinson, D. M., Maraston, C., Goddard, D., Thomas, D., \& Parikh, T., {FIREFLY (Fitting IteRativEly For Likelihood analYsis): a full spectral fitting code.} {\it Mon. Not. R. Astron. Soc.\/} {\bf 472}, 4297 (2017).
	
	\bibitem{mosdef}
	Reddy, N. A., {\it et~al.\/}, {The MOSDEF Survey: Significant evolution in the rest-frame optical emission line equivalent widths of star-forming galaxies at z=1.4-3.8.} {\it Astrophys. J.\/} {\bf 869}, 92 (2018).
		
	\bibitem{keeton10}
	Keeton, C. R., {On modeling galaxy-scale strong lens systems.} {\it Gen. Rel. Grav.\/} {\bf 42}, 2151 (2010).
	
	\bibitem{mccullykeetonwong14}
	McCully, C., Keeton, C. R., Wong, K. C., \& Zabludoff, A. I. {A new hybrid framework to efficiently model lines of sight to gravitational lenses.} {\it Mon. Not. R. Astron. Soc.\/} {\bf 443}, 3631 (2014).
	
	\bibitem{ammonswongzabludoff14}
	Ammons, S. M., Wong, K. C., Zabludoff, A. I., \& Keeton, C. R., {Mapping compound cosmic telescopes containing multiple projected cluster-scale halos.} {\it The Astronomical Journal\/} {\bf 781}, 2 (2014).
	
	\bibitem{johnsonsharonbayliss14}
    Johnson, T. L., {\it et~al.\/}, {Lens models and magnification maps of the six Hubble Frontier Fields clusters.} {\it The Astronomical Journal\/} {\bf 797}, 48 (2014).
	
	\bibitem{jullokneiblimousin07}
	Jullo, E., {\it et~al.\/}, {A Bayesian approach to strong lensing modelling of galaxy clusters.} {\it New J. Phys.\/} {\bf 9}, 447 (2007).
	
	\bibitem{faberjackson}
	Faber, S. M. \& Jackson, R. E., {Velocity dispersions and mass-to-light ratios for elliptical galaxies.} {\it Astrophys. J.} {\bf 204}, 668-683 (1976).
	
	\bibitem{witt2000}
	Witt, H. J. \& Mao, S., {On the magnification relations in quadruple lenses: a moment approach.} {\it Mon. Not. R. Astron. Soc.\/} {\bf 311}, 689-697 (2000).
	
	\bibitem{droutchornocksoderberg14}
	Drout, M. R., {\it et~al.\/}, {Rapidly evolving and luminous transients from Pan-STARRS1.} {\it Astrophys. J.\/} {\bf 794}, 23 (2014).
	
	\bibitem{hophinneyravi19}
	Ho, A. Y. Q., {\it et~al.\/}, {AT2018cow: A luminous millimeter transient.} {\it Astrophys. J.\/} {\bf 871}, 73 (2019).
	
	\bibitem{marguttimetzgerchornock19}
	Margutti, R., {\it et~al.\/}, {An embedded X-Ray source shines through the aspherical AT2018cow: Revealing the inner workings of the most luminous fast-evolving optical transients.} {\it Astrophys. J.\/} {\bf 872}, 18 (2019).
	
	\bibitem{pysynphot}
	Astrolib PySynphot (pysynphot) https://pysynphot.readthedocs.io/en/latest/ (2019).
	
	\bibitem{emcee}
	Foreman-Mackey, D., Hogg, D. W., Lang, D., \& Goodman, J., {emcee: The MCMC Hammer.} {\it Publications of the Astronomical Society of the Pacific\/} {\bf 125}, 306 (2013).
	
	\bibitem{supp_repo}
    Chen, W., {Additional Data for "Shock cooling of a red-supergiant supernova at redshift 3 in lensed images".} https://doi.org/10.5281/zenodo.6725770 (2022)
    
    \bibitem{fast}
	Kriek, M., {\it et~al.\/}, {An ultra-deep near-infrared spectrum of a compact quiescent galaxy at z = 2.2.} {\it Astrophys. J.\/} {\bf 700}, 221 (2009).
	
	\bibitem{fastpp}
	Schreiber, C. \& Dickinson, H., {FAST++.} v1.3.1 https://github.com/cschreib/fastpp (2021).
	
	\bibitem{morishitaabramsontreu17}
	Morishita, T., {\it et~al.\/}, {Characterizing intracluster light in the Hubble Frontier Fields.} {\it Astrophys. J.\/} {\bf 846}, 139 (2017).
	
	\bibitem{diegokaiserbroadhurst18}
	Diego, J. M., {\it et~al.\/}, {Dark matter under the microscope: Constraining compact dark matter with caustic crossing events.} {\it Astrophys. J.\/} {\bf 857}, 25 (2018).
	
	\bibitem{sn1987a}
	Menzies, J. W., {\it et~al.\/}, {Spectroscopic and photometric observations of SN 1987a: the first 50 days.} {\it Mon. Not. R. Astron. Soc.\/} {\bf 227}, 39 (1987).
	
	\bibitem{sn2021yja}
	Hosseinzaden, G., {\it et~al.\/}, {Weak mass loss from the red supergiant progenitor of the yype II SN 2021yja.} Preprint at arXiv:2203.08155 (2022).
	
	\bibitem{sn2013fs}
	Bullivant, C., {\it et~al.\/}, {SN 2013fs and SN 2013fr: exploring the circumstellar-material diversity in Type II supernovae.} {\it Mon. Not. R. Astron. Soc.} {\bf 476}, 1497-1518 (2018).
	
	\bibitem{valenti2016}
	Valenti, S., {\it et~al.\/}, {The diversity of Type II supernova versus the similarity in their progenitors.} {\it Mon. Not. R. Astron. Soc.} {\bf 459}, 3939–3962 (2016).
	
	\bibitem{sn2020bvc}
	Ho, A. Y. Q., {\it et~al.\/}, {SN 2020bvc: a broad-line type Ic supernova with a double-peaked optical light curve and a luminous X-ray and radio counterpart.} {\it Astrophys. J.\/} {\bf 902}, 86 (2020).
	
	\bibitem{sn2018gv}
	Yang, Y., {\it et~al.\/}, {The young and nearby normal type Ia supernova 2018gv: UV-optical observations and the earliest spectropolarimetry.} {\it Astrophys. J.\/} {\bf 902}, 46 (2020).
	
	\bibitem{ho2020}
	Ho, A. Y. Q., {\it et~al.\/}, {The Koala: a Fast Blue Optical Transient with luminous radio emission from a starburst dwarf galaxy at z = 0.27.} {\it Astrophys. J.\/} {\bf 895}, 49 (2020).
	
	\bibitem{sncosmo}
	Barbary, K., {SNCosmo: a Python library for supernova cosmology.}
	
	https://sncosmo.readthedocs.io/en/stable/index.html (2013).
	
	\bibitem{richardson2014}
	Richardson, D., Jenkins, R. L. III, Wright, J., \& Maddox, L., {Absolute-magnitude distributions of supernovae.} {\it Astrophys. J.\/} {\bf 147}, 118 (2014).
	
	\bibitem{hst_etc}
	{Hubble Space Telescope Exposure Time Calculator.} https://etc.stsci.edu (2021).
	
	\bibitem{taylor2014}
	Taylor, M., {\it et~al.\/}, {The core collapse supernova rate from the SDSS-II supernova survey.} {\it Astrophys. J.\/} {\bf 792}, 135 (2014).
	
	\bibitem{hatano1998}
	Hatano, K., Branch, D., \& Deaton, J., {Extinction and radial distribution of supernova properties in their parent galaxies.} {\it Astrophys. J.\/} {\bf 502}, 177 (1998).
	
	\bibitem{calzetti00}
	Calzetti, D., {\it et~al.\/}, {The dust content and opacity of actively star-forming galaxies.} {\it Astrophys. J.\/} {\bf 533}, 682-695 (2000).
	
	\bibitem{jeffreys_prior}
	Jeffreys, H., {An invariant form for the prior probability in estimation problems.} {\it Proceedings of the Royal Society of London. Series A, Mathematical and Physical Sciences.\/} {\bf 186}(1007) 453–461 (1946).
	
	\bibitem{kelly2016}
	Kelly, P. L., {\it et~al.\/}, {SN Refsdal: Classification as a luminous and blue SN 1987A-like Type II supernova.} {\it Astrophys. J.\/} {\bf 831}, 205 (2016).
	
	\bibitem{pastorello2012}
	Pastorello, A., {\it et~al.\/}, {SN 2009E: a faint clone of SN 1987A.} {\it Astron. Astrophys.\/} {\bf 537}, A141 (2012).
	
	\bibitem{taddia2016}
	Taddia, F., {\it et~al.\/}, {Long-rising Type II supernovae from Palomar Transient Factory and Caltech Core-Collapse Project.} {\it Astron. Astrophys.\/} {\bf 588}, A5 (2016).

    \bibitem{coe10}
    Coe, D., Ben{\'\i}tez, N., Broadhurst, T., \& Moustakas, L. A., A high-resolution mass map of galaxy cluster substructure: LensPerfect analysis of A1689. {\it Astrophys. J.\/} {\bf 723}, 1678 (2010).
	
\end{thebibliography}


\begin{addendum}
 \item[Data availability:] The {\it HST} data used for this study can be retrieved from the NASA Mikulski Archive for Space Telescopes (http://archive.stsci.edu). The supernova is found in the {\it HST} imaging of the Abell 370 field acquired from program GO-11591 (PI J.-P. Kneib). The LBT spectroscopy data are available from the LBT archive (http://archive.lbto.org). The Keck MOSFIRE data can be retrieved from the Keck Observatory Archive (https://www2. keck.hawaii.edu/koa/public/koa.php). The HFF data and models can be downloaded from https://archive. stsci.edu/prepds/frontier/lensmodels/\#modelsandinput. Additional data including the MMT spectroscopic data, the {\it HST} coaddition and image differencing data, the {\tt GALFIT} scripts and resulting models, the {\it HST} photometry data of the SN host galaxy, the SN light-curve fitting script and resulting MCMC data, and our best-fit {\tt GLAFIC} lens model are available from https://doi.org/10.5281/zenodo.6725770.
 \item[Acknowledgments:] This work was supported by the {\it HST} Cycle 27 Archival Research program (grant AR-15791), as well as by GO-15936 and GO-16278.
 We utilise gravitational lensing models produced by 
 the GLAFIC group. The lens modelling was partially funded by the {\it HST} Frontier Fields program conducted by STScI. The lens models were obtained from the Mikulski Archive for Space Telescopes.
 Some of the observations reported here were obtained at the MMT Observatory, a joint facility of the Smithsonian Institution and the University of Arizona.
 The data presented here were obtained in part at the LBT Observatory. The LBT is an international collaboration among institutions in the United States, Italy, and Germany. 
 Some of the data presented herein were obtained at the W.~M. Keck Observatory, which is operated as a scientific partnership among the California Institute of Technology, the University of California, and the National Aeronautics and Space Administration (NASA). The Observatory was made possible by the generous financial support of the W.~M. Keck Foundation. The authors wish to recognise and acknowledge the very significant cultural role and reverence that the summit of Maunakea has always had within the indigenous Hawaiian community.  We are most fortunate to have the opportunity to conduct observations from this mountain.
 P.L.K. is supported by United States National Science Foundation (NSF) grant AST-1908823.
 J.M.D. acknowledges support from projects PGC2018-101814-B-100 and MDM-2017-0765.
 M.O. acknowledges support from World Premier International Research Center Initiative, MEXT, Japan, and JSPS KAKENHI grants JP20H00181, JP20H05856, JP22H01260, and JP18K03693.
 A.Z. acknowledges support by grant 2020750 from the United States-Israel Binational Science Foundation (BSF) and grant 2109066 from the U.S. NSF, and by the Ministry of Science \& Technology, Israel.
 A.V.F. is grateful for assistance from the Christopher R. Redlich Fund, the U.C. Berkeley Miller Institute for Basic Research in Science (where he was a Miller Senior Fellow), and many individual donors.
 We acknowledge Dr. WeiKang Zheng's help with the Keck MOSFIRE observations.
 \item[Author contributions:]
 W.C. analysed the {\it HST}, Keck, LBT, and MMT data, wrote the manuscript, and developed simulations.
 P.L.K. aided the interpretation of the events, planned the Keck, LBT, and MMT observations, edited the manuscript, and cross-checked the results.
 M.O. constructed the gravitational-lensing model for this work. 
 T.J.B., J.M.D., and A.Z. helped with the {\it HST} proposal and reviewed the manuscript.
 P.L.K., N.E., and A.Z. acquired the Keck observation. P.L.K. and N.E acquired the LBT observation.
 A.Z. constructed independent lens models for the cluster to verify the time-delay results.
 A.V.F. obtained the Keck time, helped acquire the Keck MOSFIRE observations, and edited the {\it HST} proposal and this manuscript.
 T.T. contributed to the manuscript and interpretation.
 \item[Competing Interests:] The authors declare no competing financial interests.
 
 \item[Additional information:]
 \item[]Correspondence and requests for materials should be addressed to Wenlei Chen~(email: chen6339@umn.edu).
 \item[]Reprints and permissions information is available at www.nature.com/reprints.
\end{addendum}

\clearpage
\thispagestyle{empty}
\begin{figure}
	\centering
	\includegraphics[angle=0,width=5.2in]{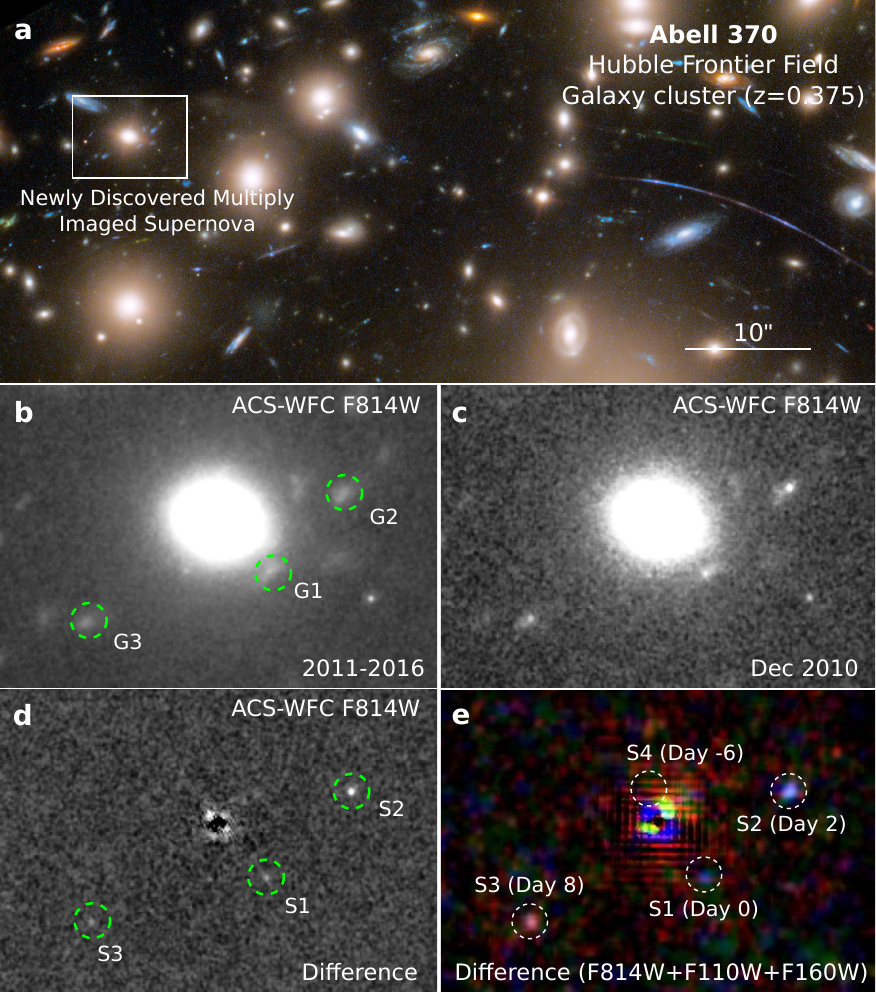}
	\caption{\linespread{1.0}\selectfont{}A multiply imaged SN discovered in the archival {\it HST} imaging. {\bf a}. Newly discovered multiply imaged SN in a gravitationally lensed galaxy at $z\approx3$ in the Abell~370 galaxy-cluster field. {\bf b}. An example deep template ACS-WFC F814W image of the field, where three images of the lensed galaxy or dwarf galaxy are marked by green dashed circles labelled as G1, G2, and G3. {\bf c}. Image of the newly identified event in 2010 December from the same bandpass filter. {\bf d}. The difference image where three SN images are marked as S1, S2, and S3 by green dashed circles. We have named the images according to the order in which the observed light was emitted, as predicted by our gravitational-lens model. {\bf e}. Mixed difference imaging from the ACS-WFC F814W, WFC3-IR F110W, and WFC3-IR F160W bandpasses. The de-noised difference images from F160W, F110W, and F814W are assigned to the red, green, and blue channels, respectively, providing a mixed image with pseudocolour. White dashed circles mark the predicted positions from our best-fit lens model. Image in {\bf a} by NASA, ESA/Hubble, HST Frontier Fields.}
\end{figure}

\clearpage
\thispagestyle{empty}
\begin{figure}
	\centering
	\includegraphics[angle=0,width=4.5in]{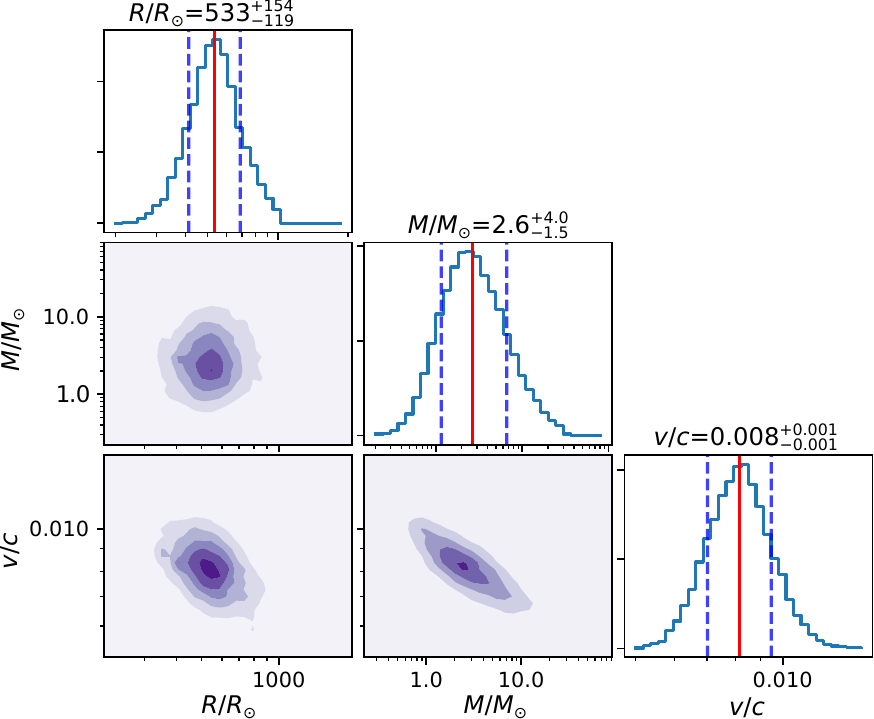}
	\caption{\linespread{1.0}\selectfont{}Posterior distributions of the progenitor radius $R$, envelope mass $M$, and shock velocity $v$ from the RSG model for $R_V=3.1$. The red solid and the blue dashed lines overplotted on each histogram denote the median and the 68\% confidence interval of each distribution, respectively. From the posterior distributions, we constrained the radius of the progenitor RSG to $533^{+154}_{-119}\,R_\odot$ and the envelope mass to $2.6^{+4.0}_{-1.5}\,M_\odot$. Distributions of all free parameters are shown in Extended Data Fig.\,4.}
\end{figure}

\clearpage
\thispagestyle{empty}
\begin{figure}
	\centering
	\includegraphics[angle=0,width=4.2in]{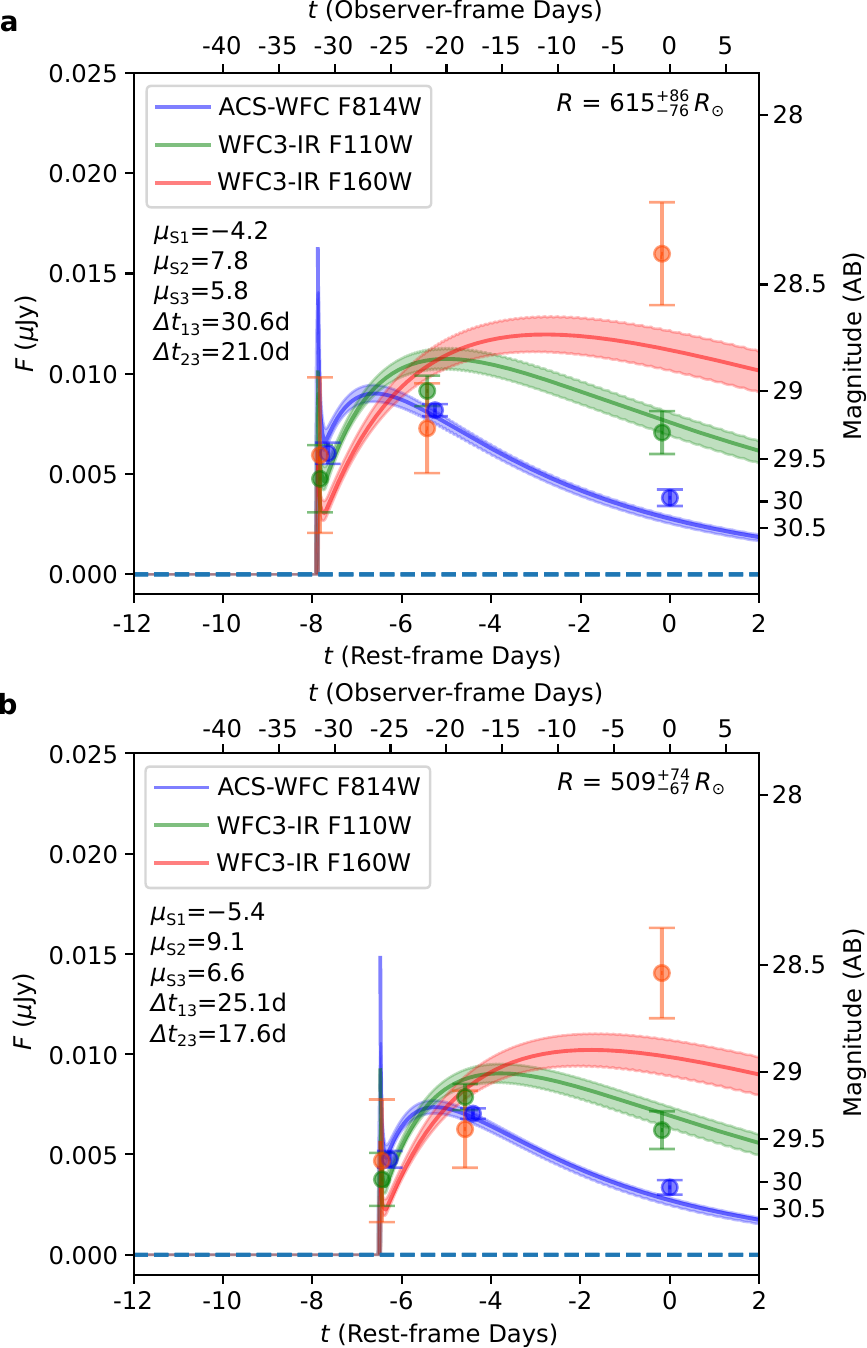}
	\caption{\linespread{1.0}\selectfont{}Two examples of the reconstructed light curve of the SN with fixed time delays and magnifications. {\bf a}. Reconstructed light curves from the RSG model for $R_V=3.1$ where the magnifications and time delays are set to the best-fit values from the lens model as listed in Extended Data Table\,1b, where $F$ and $t$ are respectively the reconstructed flux density and time. Solid lines show the best-fit light curves and shaded regions are the 68\% confidence intervals of the flux density. The radius of the progenitor is constrained to $615^{+86}_{-76}$\,$R_\odot$. {\bf b}. Reconstructed light curves from the RSG model for $R_V=3.1$, while we chose the magnifications to be $1\sigma$ (standard deviation) larger, and the time delays to be $1\sigma$ smaller than their best-fit values. This infers a smaller progenitor radius as $509^{+74}_{-67}$\,$R_\odot$ compared to the result using best-fit magnifications and time delays.}
\end{figure}

\clearpage
\thispagestyle{empty}
\begin{figure}
	\centering
	\includegraphics[angle=0,width=4.8in]{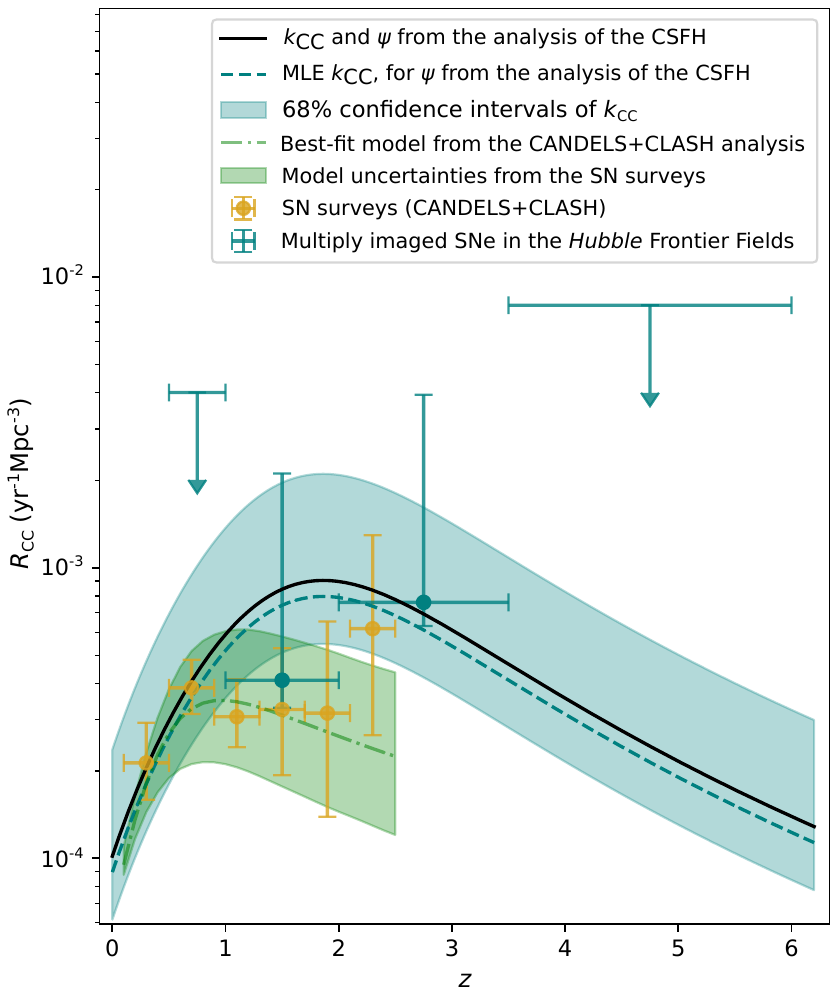}
	\caption{\linespread{1.0}\selectfont{}Volumetric CCSN rate ($R_\mathrm{CC}$) as a function of redshift. Black solid curve is from the star-formation rate density ($\psi$)\cite{MD14} and $k_\mathrm{CC}=0.0068~M_\odot^{-1}$ for a Salpeter IMF\cite{salpeter_imf}. Dark cyan data points are the MLE $R_\mathrm{CC}$ values estimated from the detection of multiply imaged SNe in the six {\it Hubble} Frontier Fields in the last decade. Horizontal error bars show the redshift bins, and vertical error bars give the 68\% confidence intervals. Cyan shaded region is the range of the estimated $R_\mathrm{CC}$ based on $\psi$ from the analysis of the CSFH\cite{MD14} and the 68\% confidence interval of $k_\mathrm{CC}$ from the detection of multiply imaged SNe. Golden data points, dash-dotted green line, and green shaded region are $R_\mathrm{CC}$ constraints from SN surveys, the best-fit star-formation model ($\psi$ and $k_\mathrm{CC}$), and the error region of the model from the CANDELS+CLASH analysis\cite{strolger15}.}
\end{figure}

\clearpage
\section*{Extended Data}
\thispagestyle{empty}

\begin{figure}
	\centering
	\includegraphics[angle=0,width=6.8in]{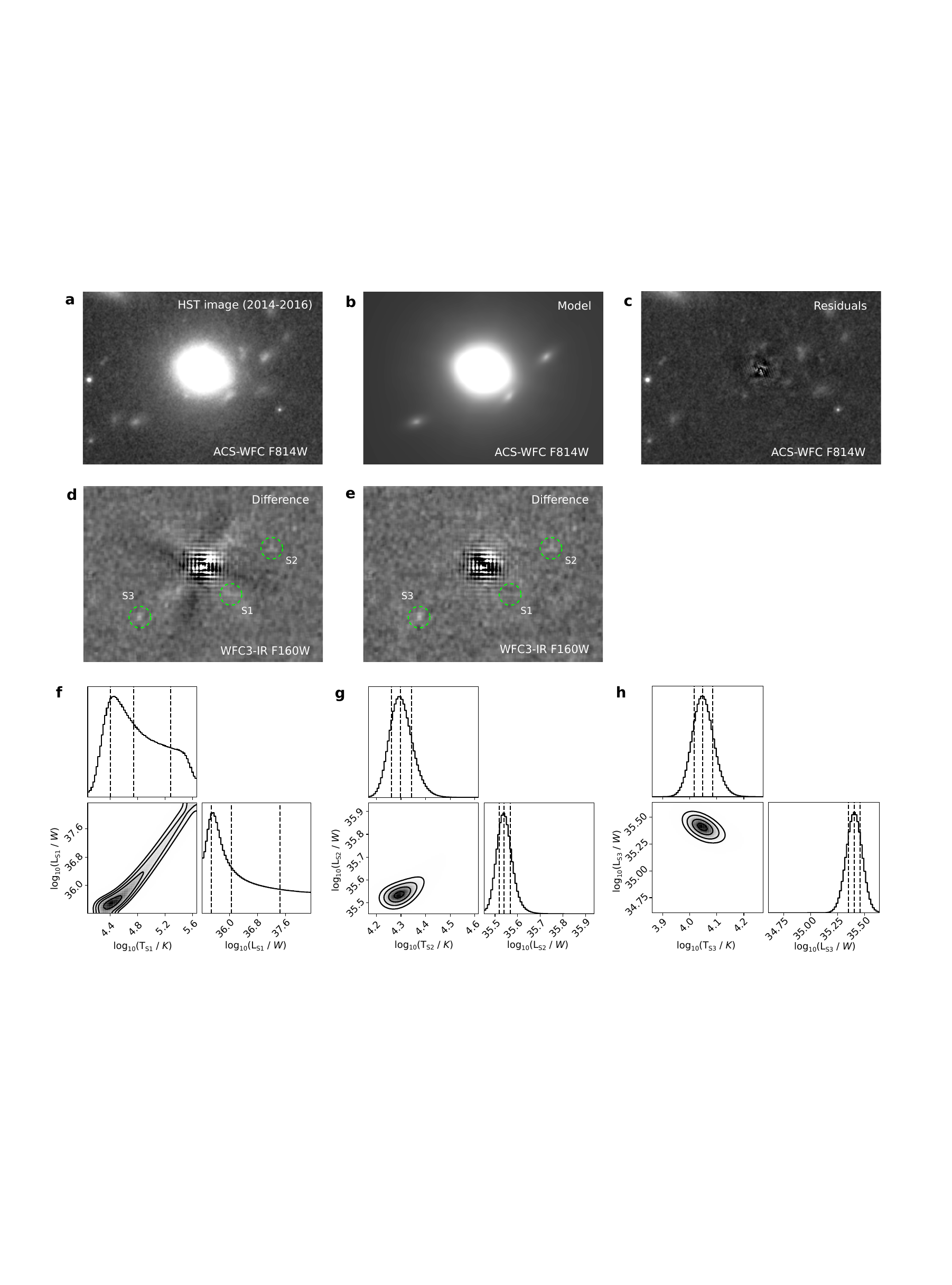}
\end{figure}
\clearpage
\thispagestyle{empty}
\subsection{Extended Data Fig.\,1}
Photometry of the multiply imaged SN: Image differencing and the black-body fitting. We used {\tt GALFIT}\cite{galfit} to fit bright sources in the lensing system and subtracted the best-fit model from the field prior to measuring the flux. Panel\,{\bf a}--panel\,{\bf c} show the template image, the best-fit {\tt GALFIT} model, and the residual through the {\it HST} ACS-WFC F814W filter. Panel\,{\bf d} shows the WFC3-IR F160W difference image by subtracting the template from the event image directly. Panel\,{\bf e} shows the difference image using our {\tt GALFIT}-based method, where we subtracted the PSF-convolved best-fit {\tt GALFIT} source models from the images, and then calculated the difference from {\tt GALFIT} residuals between the coadded images and the event images. We can see that this method can reduce the significant residual from the mismatched PSFs around bright sources as shown in panel\,{\bf d}. Panel\,{\bf f}--panel\,{\bf h} show distributions of effective temperature and luminosity from the MCMC samples from fitting the blackbody emission into the photometry of each SN image.

\begin{figure}
	\centering
	\includegraphics[angle=0,width=6.8in]{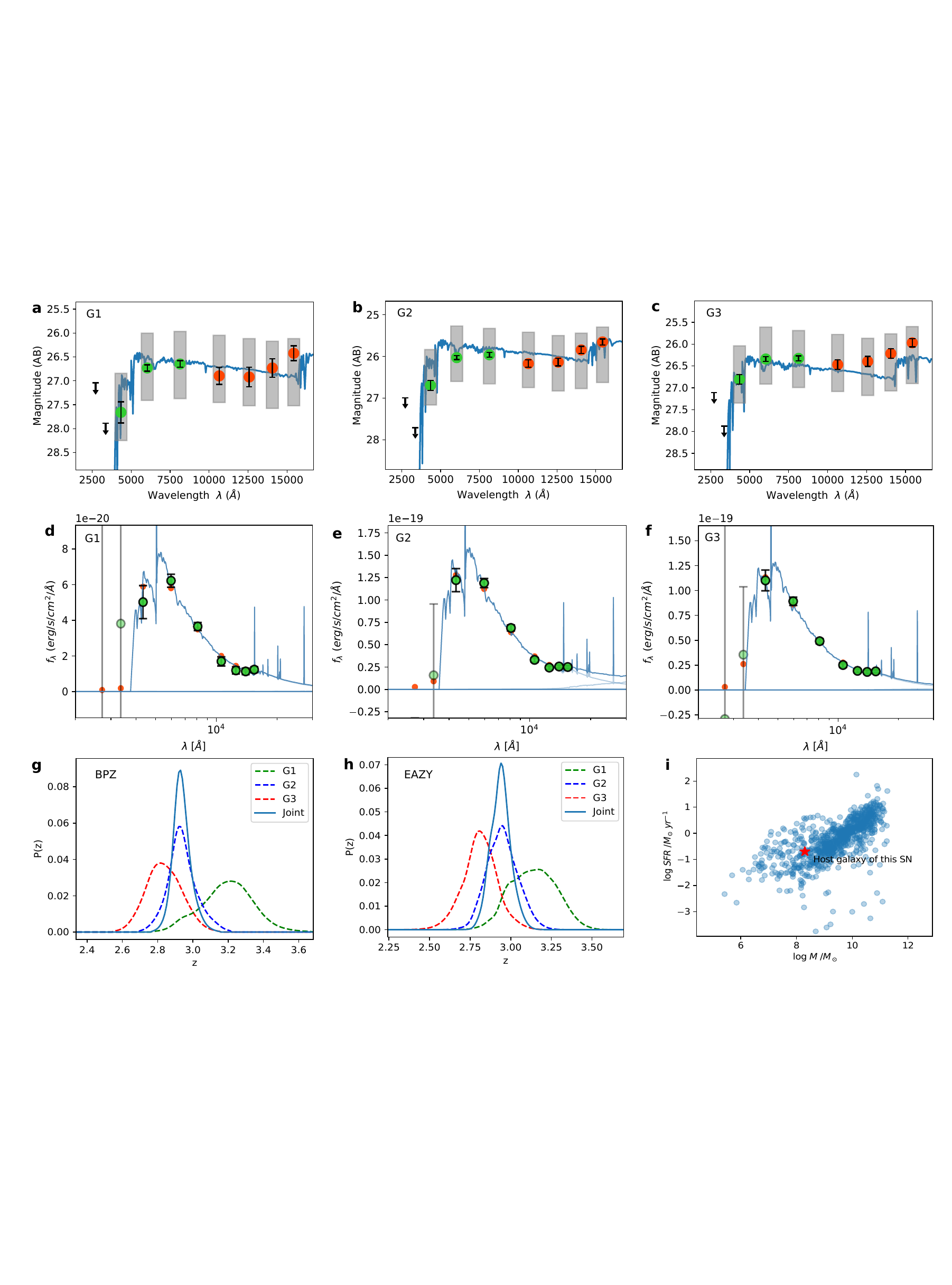}
\end{figure}
\clearpage
\thispagestyle{empty}
\subsection{Extended Data Fig.\,2}
Photometry and photometric redshift of the SN host galaxy. Panel\,{\bf a}--panel\,{\bf c} show broadband spectral energy distributions (SEDs) and best-fitting {\tt BPZ}\cite{bpz1} templates for G1, G2, and G3. The coloured circles designate the observed AB magnitudes with uncertainties measured from ACS-WFC (green) and WFC3-IR (red) images, while arrows correspond to 95\% upper limits. Dark-blue curves plot the best-fit spectral templates. Grey rectangles mark the magnitudes calculated from the best-fitting {\tt BPZ} templates (with approximate uncertainties) for those filters\cite{coe10}. Panel\,{\bf d}--panel\,{\bf f} show broadband SEDs and best-fitting {\tt EAZY}\cite{bpz1} templates for G1, G2, and G3. The data points are the observed flux with uncertainties. Blue curves show the best-fit spectral template. The three panels on the right display the posterior probability distribution of the photometric redshift. Panel\,{\bf g} and panel\,{\bf h} show photometric redshift probability distributions for G1 (green-dash line), G2 (blue-dash line), G3 (red-dash), and the joint analysis (solid line) derived from {\tt BPZ} and {\tt EAZY} fitting. Panel {\bf i} shows the distribution of CCSN host-galaxy stellar mass ($\mathrm{M}$) and star-formation rate ($\mathrm{SFR}$)\cite{schulze2021}, where the red star marks $\mathrm{M}$ and $\mathrm{SFR}$ of the host galaxy of the newly discovered SN.

\begin{figure}
	\centering
	\includegraphics[angle=0,width=6.8in]{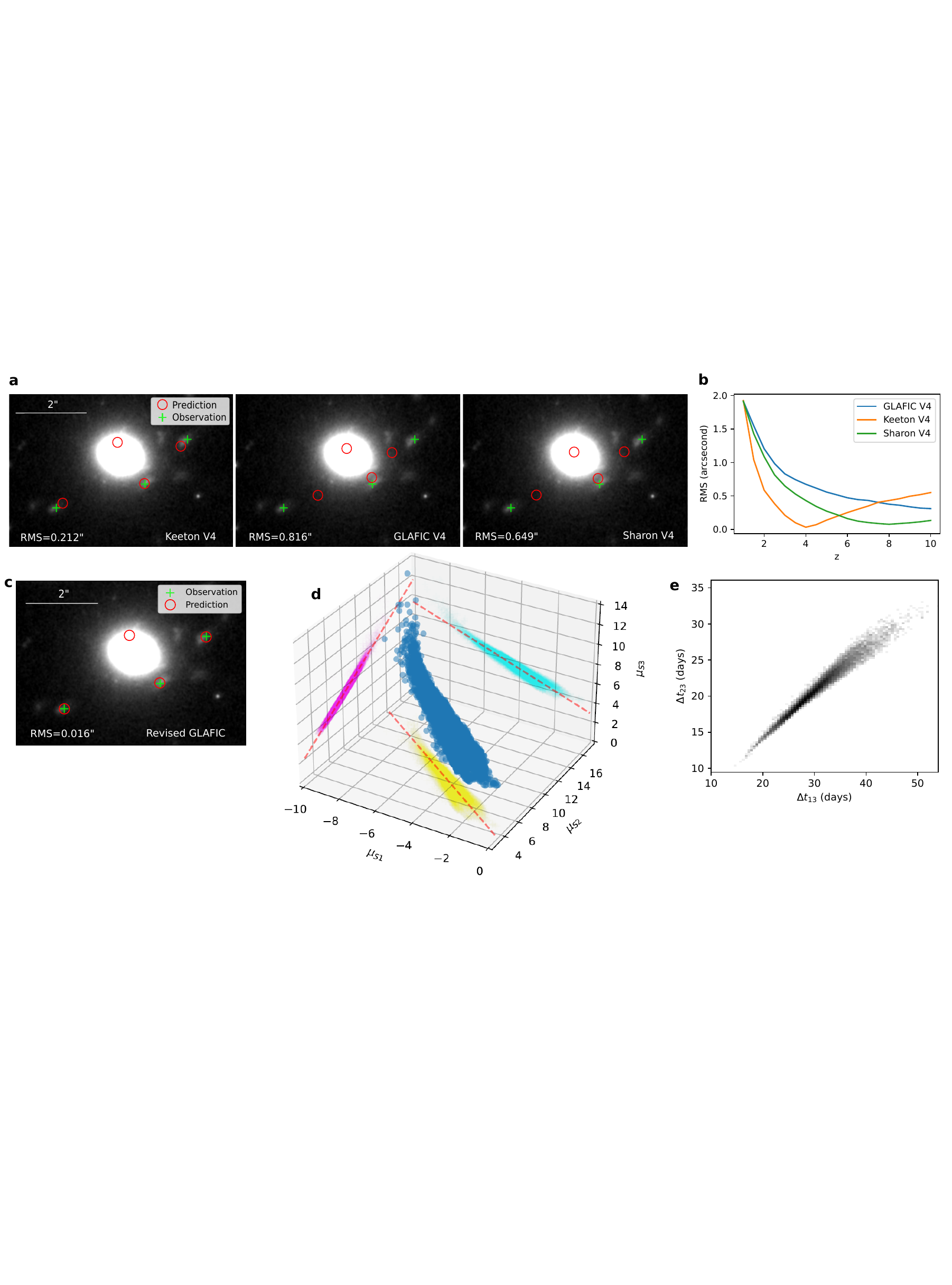}
\end{figure}
\clearpage
\thispagestyle{empty}
\subsection{Extended Data Fig.\,3}
Modelling the strongly lensed images. {\bf a}. Best-fit multiple-image geometry from the published Keeton V4, GLAFIC V4, and Sharon V4 models, respectively, by minimising the RMS angular separation of the predicted images from the real images in the image plane, for $z=3$. {\bf b}. Best-fit RMS angular separation of the predicted images as a function of $z$ from the published Keeton V4, GLAFIC V4, and Sharon V4 models. {\bf c}. Best-fit multiple-image geometry from our revised {\tt GLAFIC} model. {\bf d}. Distribution of $\mu_{S1}$, $\mu_{S2}$, and $\mu_{S3}$ from the MCMC samples given by our revised {\tt GLAFIC} model, where $\mu_1$, $\mu_2$, and $\mu_3$ are predicted magnifications of images S1, S2, and S3 (respectively). Cyan, yellow, and magenta markers show the distribution projected on the $\mu_{S1}-\mu_{S3}$, $\mu_{S1}-\mu_{S2}$, and $\mu_{S2}-\mu_{S3}$ planes, where red-dashed lines are the best-fit lines from linear regressions on the projected distribution. {\bf e}. Correlations between $\Delta t_{13}$ and $\Delta t_{23}$ from the MCMC samples, where $\Delta t_{13}$ and $\Delta t_{23}$ are the time delay of S1 and S2 relative to S3, respectively.

\begin{figure}
	\centering
	\includegraphics[angle=0,width=6.8in]{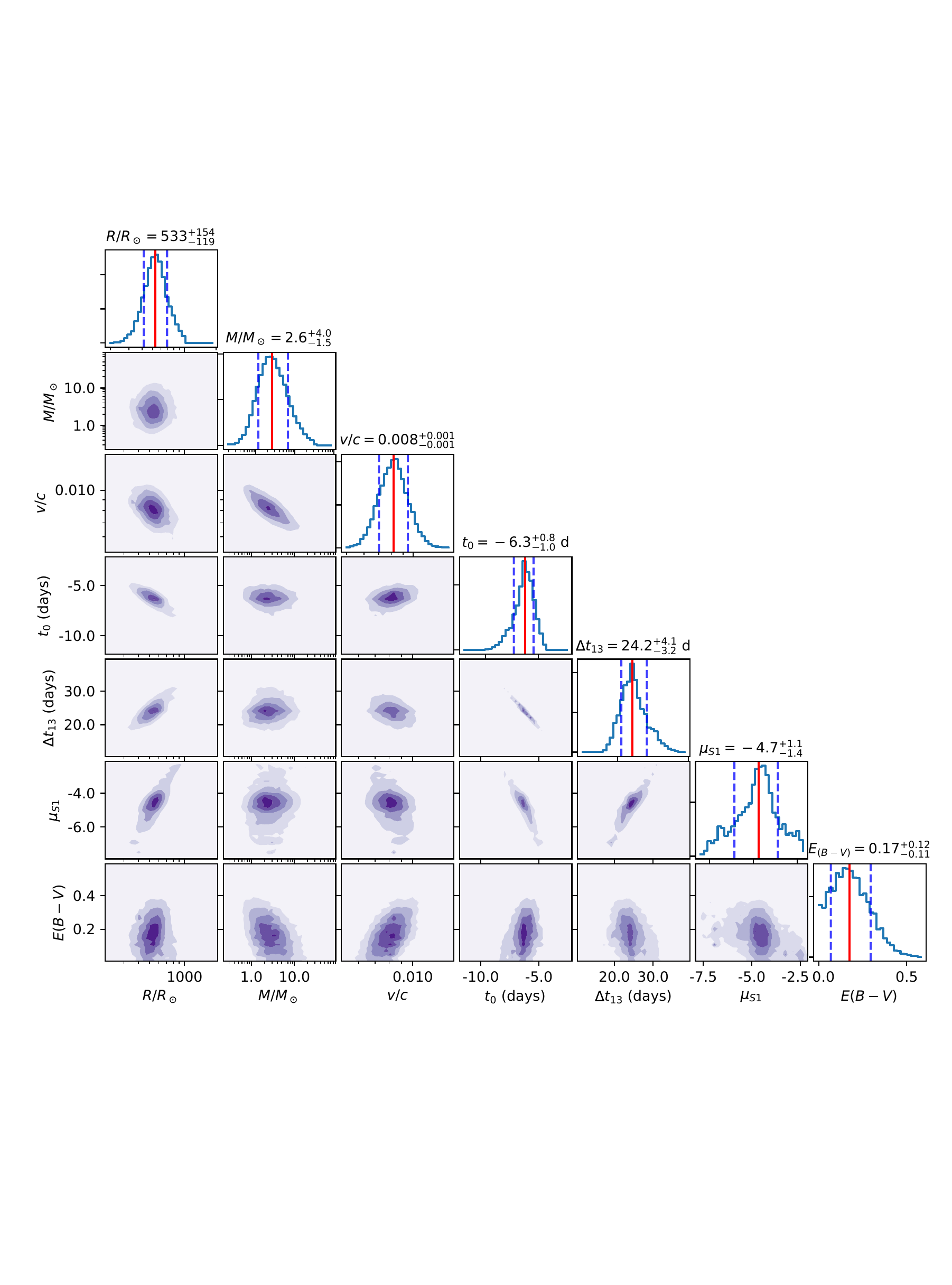}
\end{figure}
\clearpage
\thispagestyle{empty}
\subsection{Extended Data Fig.\,4}
Distribution of parameters from fitting the RSG model to the observations. Here, $R$ is the extended envelope radius, $M$ is the envelope mass, $v$ is the shock velocity, $t_0$ is the initial time, and $\Delta t_{13}$ and $\mu_{S1}$ are respectively the time delay and magnification of the S1 image. We set $t=0$ at the observation time of the S3 image in the ACS-WFC F814W band. The time delay of each image is defined as the difference of its light-travel time from that of S3. The result is for $R_V=3.1$.

\begin{figure}
	\centering
	\includegraphics[angle=0,width=6.8in]{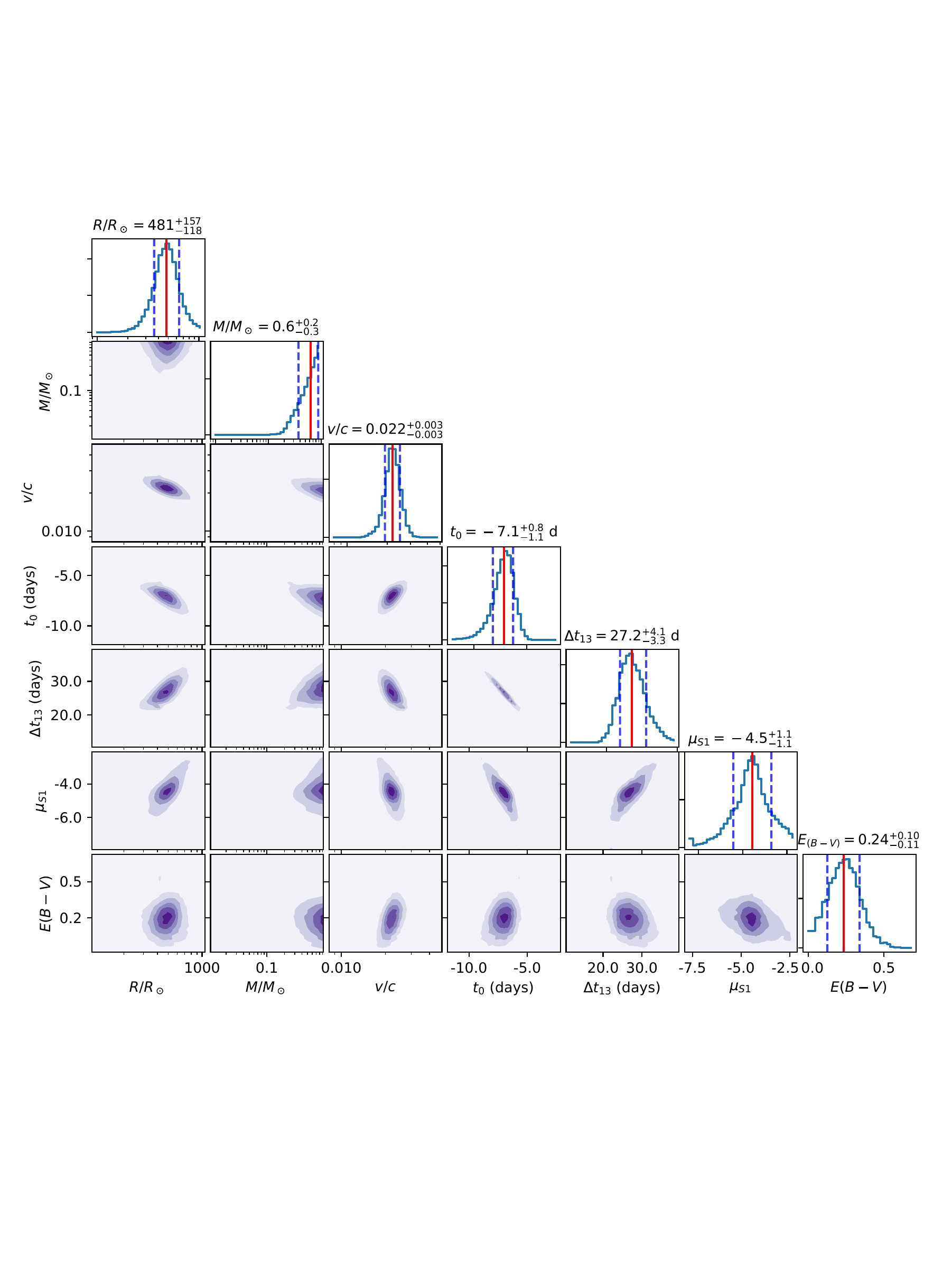}
\end{figure}
\clearpage
\thispagestyle{empty}
\subsection{Extended Data Fig.\,5}
Distribution of parameters from fitting the CSM-homologous model to the observations. The result is for a prior on the CSM mass of $M< 1\, M_\odot$. The parameters are the same as in Extended Data Fig.\,5. Data of the MCMC samples for all models with all our choices of $R_V$ and priors are available from an online repository\cite{supp_repo}.

\begin{figure}
	\centering
	\includegraphics[angle=0,width=6.8in]{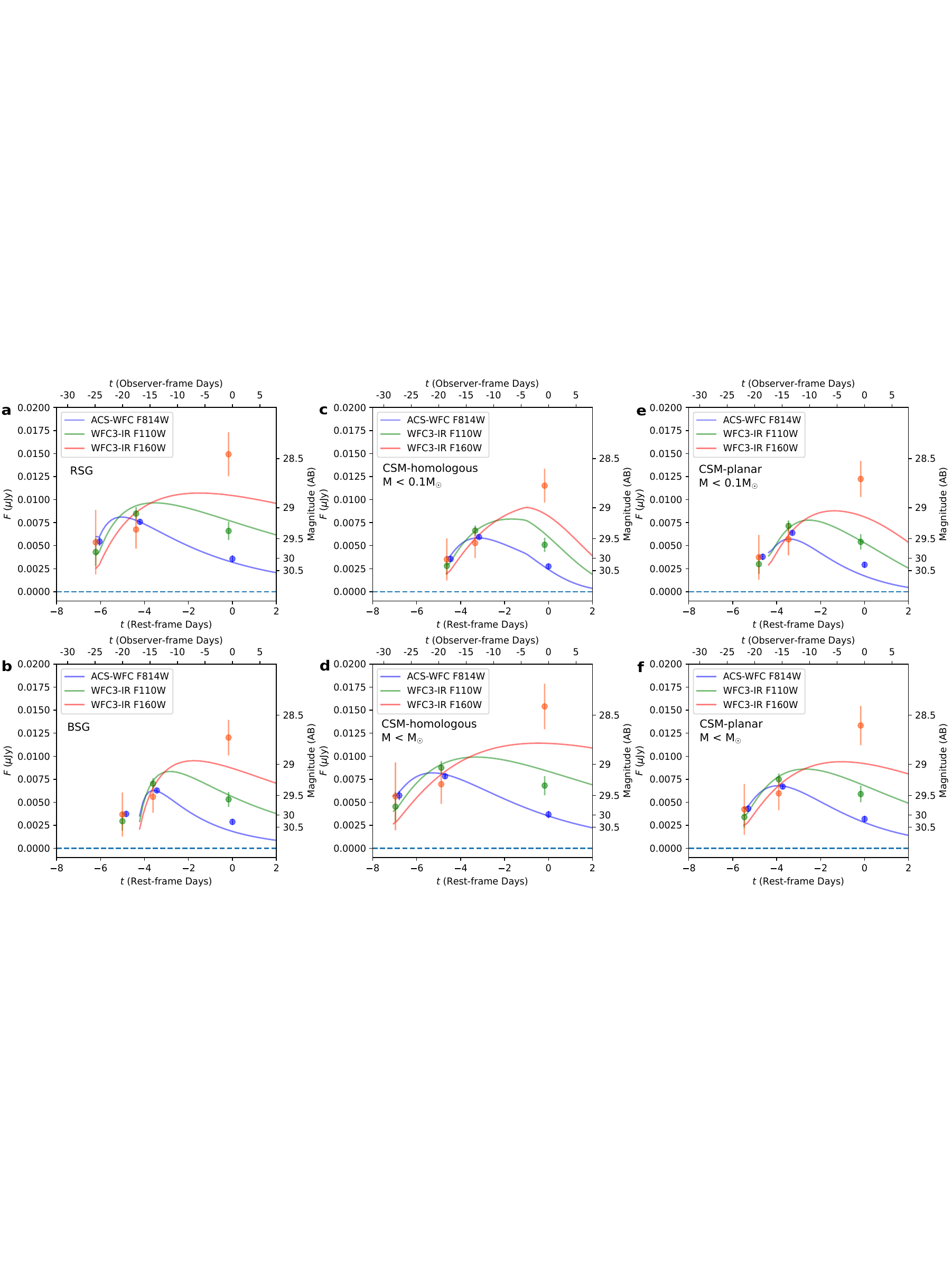}
\end{figure}
\clearpage
\thispagestyle{empty}
\subsection{Extended Data Fig.\,6}
Reconstructed light curves from the best-fit model parameters from all the light-curve models for $R_V=3.1$. Panels {\bf a}--{\bf f} are from the RSG model, the BSG model, the CSM-homologous model for a prior on the CSM mass of $M< 0.1\, M_\odot$, the CSM-homologous model for a prior on the CSM mass of $M< 1\, M_\odot$, the CSM-planar model for a prior on the CSM mass of $M< 0.1\, M_\odot$, and the CSM-planar model for a prior on the CSM mass of $M< 1\, M_\odot$, respectively. Reconstructed light curves with all our choices of $R_V$ are available from an online repository\cite{supp_repo}.

\begin{figure}
	\centering
	\includegraphics[angle=0,width=6.8in]{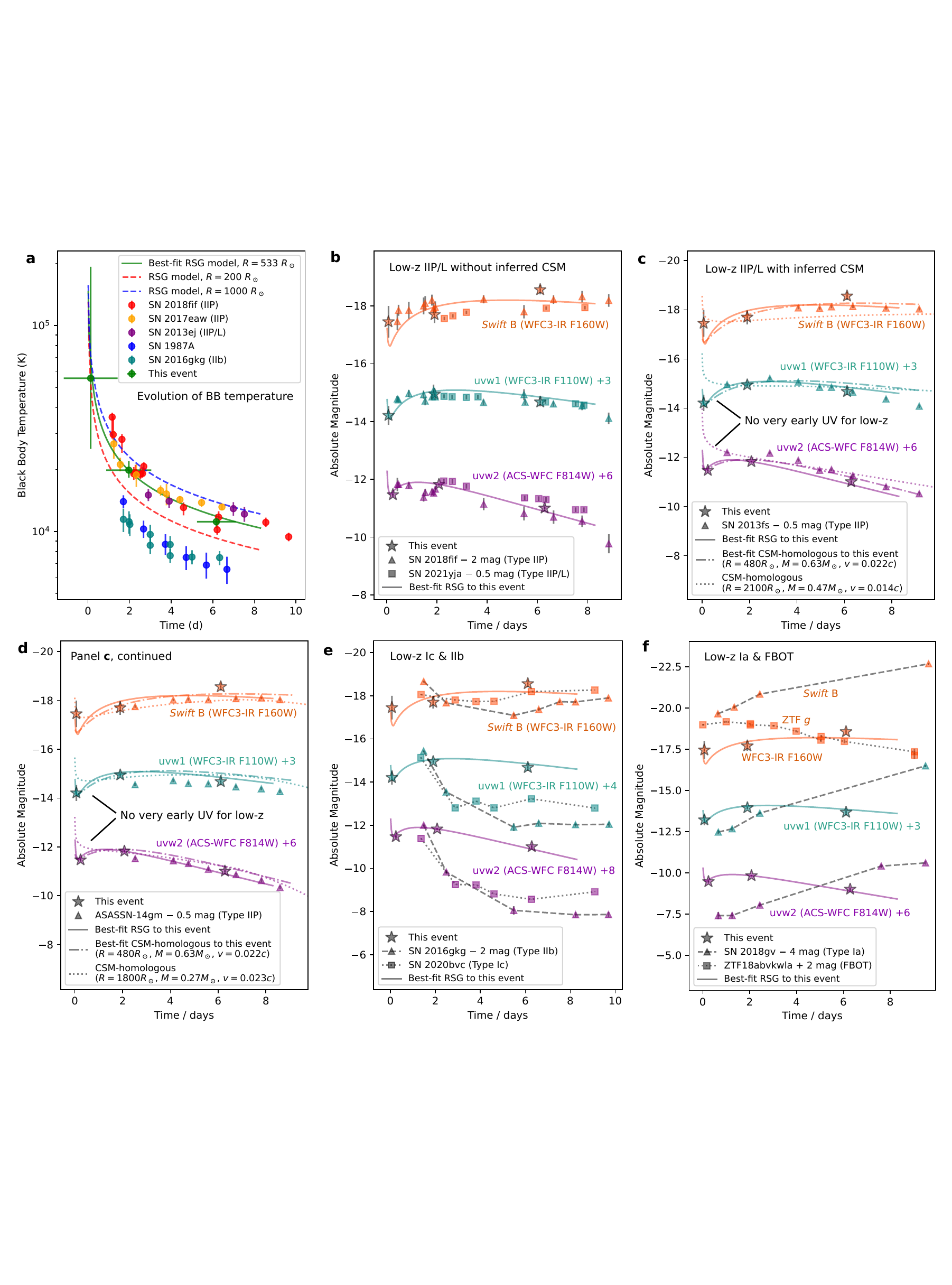}
\end{figure}
\clearpage
\thispagestyle{empty}
\subsection{Extended Data Fig.\,7}
Comparisons of the reconstructed SN light curve with early UV light curves from other SNe. Panel {\bf a} shows the early-time evolution of the effective black-body temperature of the newly discovered SN (green data points, where error bars are 68\% confidence intervals), where the effective temperatures of the SN images are obtained by independently fitting the blackbody emission into the photometry of each SN image. The green-solid line is from the best-fit RSG model. Red and blue dashed lines are two examples from the RSG model with progenitor radius of $200\,R_\odot$ and $1000\,R_\odot$, respectively, for the same envelope mass and shock velocity from our best-fit RSG model for the newly discovered SN. Red, yellow, blue, cyan data points show the early-time evolution of SN 2018fif\cite{sn2018fif} (Type IIP), SN 2013ej\cite{valenti2014} (Type IIP/L), SN 2017eaw\cite{szalai2019,rui2019} (Type IIP), SN 1987A\cite{sn1987a}, and SN 2016gkg\cite{sn2016gkg,tartaglia2017,arcavi2017,p21} (Type IIb). Stars and solid curves in panels {\bf b}--{\bf f} show the absolute magnitude of the newly discovered SN and reconstructed light curves from the best-fit RSG model, respectively, with arbitrary magnitude offsets for better visualisation. For this SN, orange, teal, and purple points and lines are for ACS-WFC F814W, WFC3-IR F110W, and WFC3-IR F160W, respectively. In the rest frame of the SN, central wavelengths of these three filters fall into the {\it Swift}-UVOT's B, UVW1, and UVW2 bands. In panel {\bf b}, we compare the SN's early light curve to the {\it Swift}-UVOT observations of SN 2018fif\cite{sn2018fif} and SN 2021yja\cite{sn2021yja}--two Type II SNe whose early light curves indicate only small amounts of CSM around their progenitors\cite{sn2018fif,sn2021yja}. Panels {\bf c} and {\bf d} show the comparison to the two Type II SNe, SN 2013fs\cite{sn2013fs} and ASASSN-14gm\cite{valenti2016}, which are believed to have dense CSM shells around their progenitors. In panels {\bf c} and {\bf d}, we plot the light curve from our best-fit CSM-homologous model for the newly discovered SN (dash-dotted lines) and light curves from the CSM-homologous model for the CSM radius and mass given by the analyses of CSM-rich CCSNe\cite{morozova2017,morozova2018} (dotted lines) for the two Type II SNe. Panels {\bf e} and {\bf f} show such comparisons to the early UV light curves of SN 2016gkg\cite{tartaglia2017} (Type IIb), SN 2020bvc\cite{sn2020bvc} (Type Ib/c), SN 2018gv\cite{sn2018gv} (Type Ia), and ZTF18abvkwla\cite{ho2020} (FBOT). ZTF-g band light curve is shown for ZTF18abvkwla at $z=0.27$, corresponding to the rest-frame wavelength of 3820\,\AA. All magnitudes are AB magnitudes.

\begin{figure}
	\centering
	\includegraphics[angle=0,width=6.8in]{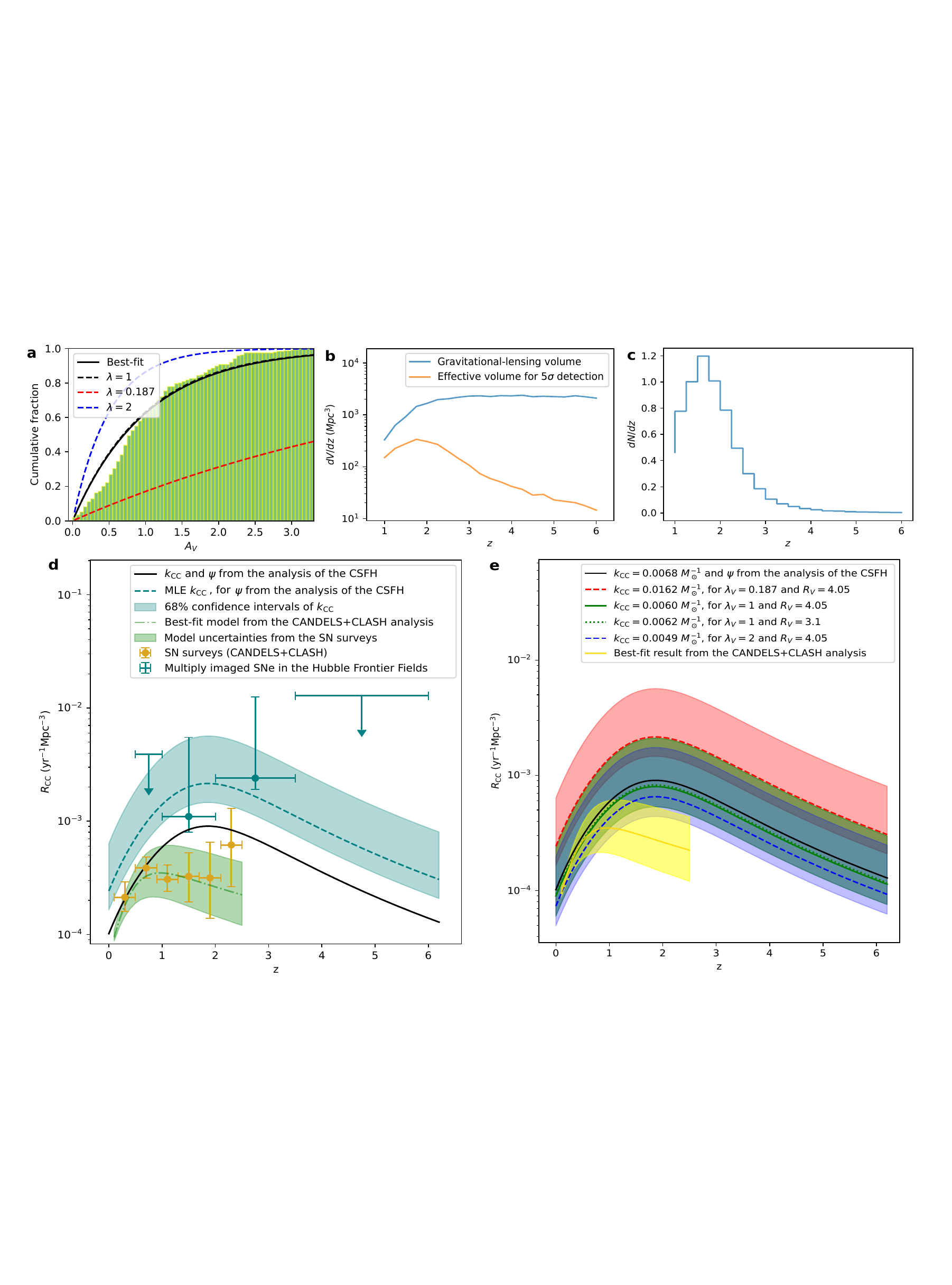}
\end{figure}
\clearpage
\thispagestyle{empty}
\subsection{Extended Data Fig.\,8}
Estimating the CCSN rate based on observations of the multiply imaged SNe. {\bf a}. Cumulative distribution of CCSN host galaxies as a function of the host-galaxy extinction $A_V$\cite{kelly2012}. The black solid line shows the cumulative distribution function of the exponential distribution $P(A_V)=\lambda_V \exp(-\lambda_V A_V)$ with best-fit $\lambda_V=0.98$. Black, red, and blue dashed lines show the cumulative distribution function of the exponential distribution with $\lambda_V=1$, $\lambda_V=0.187$, and $\lambda_V=2$, respectively. {\bf b}. Differential comoving volume ($dV/dz$) for the strongly lensed sources as a function of redshift $z$. Blue curve shows the lensed volume for all the sources in the six {\it Hubble} Frontier Fields within a $0.03^\circ\times0.03^\circ$ search window for each cluster field. Orange curve shows the effective lensing volume for sources that can be significantly ($>5\sigma$) detected by {\it HST} in the last decade. {\bf c}. Differential number of detectable multiply imaged core-collapse supernovae ($dN/dz$) above the $5\sigma$ signal-to-noise-ratio level by {\it HST} in the last decade as a function of $z$. {\bf d}. Volumetric CCSN rate ($R_\mathrm{CC}$) as a function of redshift. Contents are the same as in Fig.\,4, but for $\lambda_V=0.187$. {\bf e}. Volumetric CCSN rate ($R_\mathrm{CC}$) as a function of redshift for different choices of $\lambda_V$ and $R_V$. Red, green, blue shaded regions are for $\lambda_V=0.187$, $\lambda_V=1$, and $\lambda_V=2$, respectively, for $R_V=4.05$. Red, green, blue lines are from the MLE $k_\mathrm{CC}$ for the three $\lambda_V$ options. The green dotted line is the same as the green solid line, but for $R_V=3.1$.

\begin{figure}
	\centering
	\includegraphics[angle=0,width=6.8in]{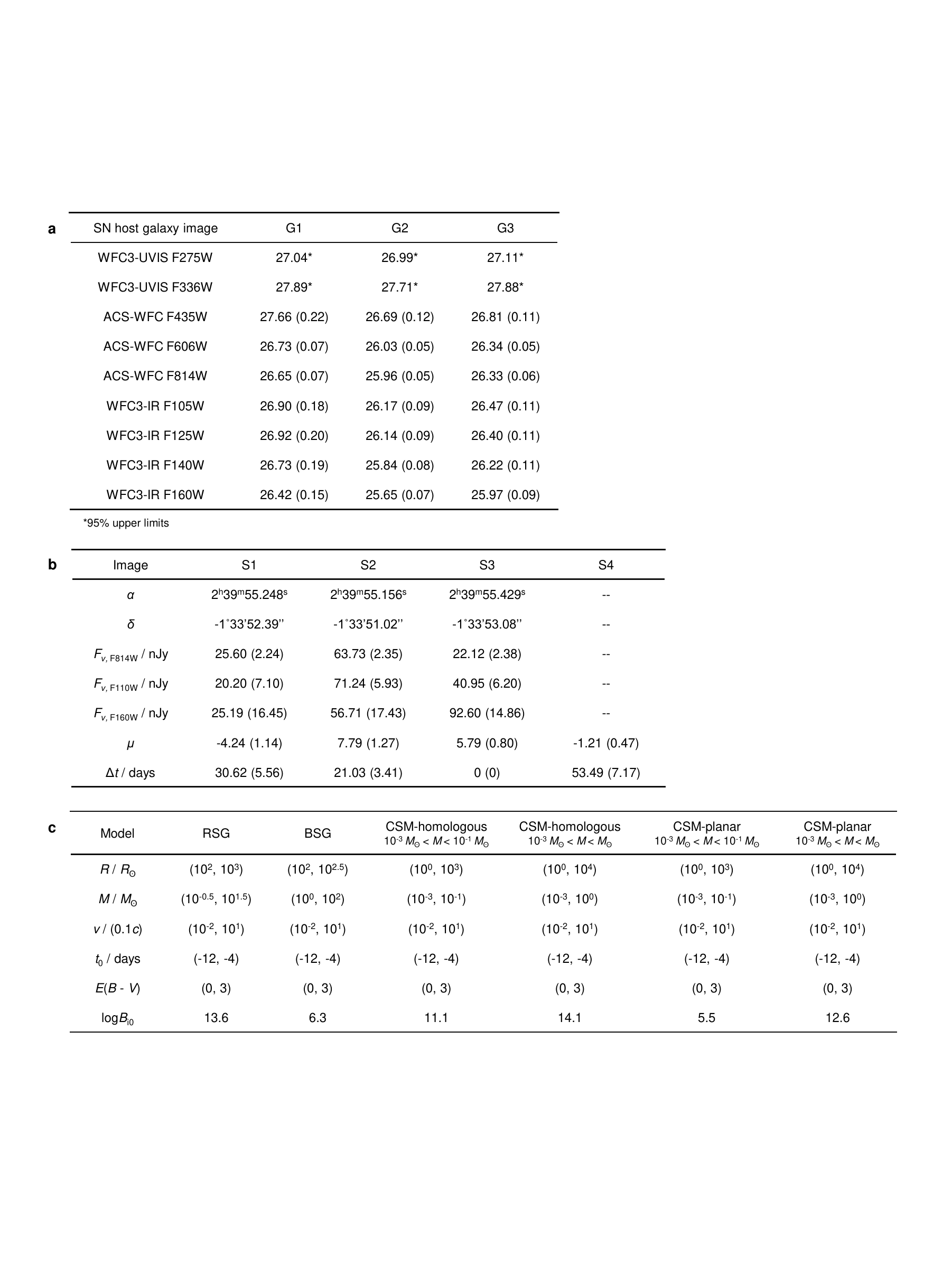}
\end{figure}
\clearpage
\thispagestyle{empty}
\subsection{Extended Data Table\,1}
Results of photometry, gravitational-lensing modelling, and the Bayesian test of light-curve models. {\bf a}. Photometry of G1, G2, and G3 in units of AB magnitude. {\bf b}. Positions ($\alpha$, $\delta$), flux density ($f$), magnification ($\mu$), and time delay ($\Delta t$) of the multiple SN images. Flux density and time delay are in units of nanoJanskys (nJy) and days, respectively, where the time delay is defined as the relative delay with respect to S3. Photometry has been corrected for foreground Milky Way extinction. A negative magnification indicates an opposite parity. Numbers in the parentheses are the standard deviations. {\bf c}. Ranges of the prior density of physical parameters $R$, $M$, $v$, $t_0$, and $E(B-V)$ for the light-curve models and logarithm Bayes factors ($\log{B_{i0}}$) of the $i$-th light-curve model against the constant-flux model (the null hypothesis) for $R_V=3.1$.

\begin{figure}
	\centering
	\includegraphics[angle=0,width=6.8in]{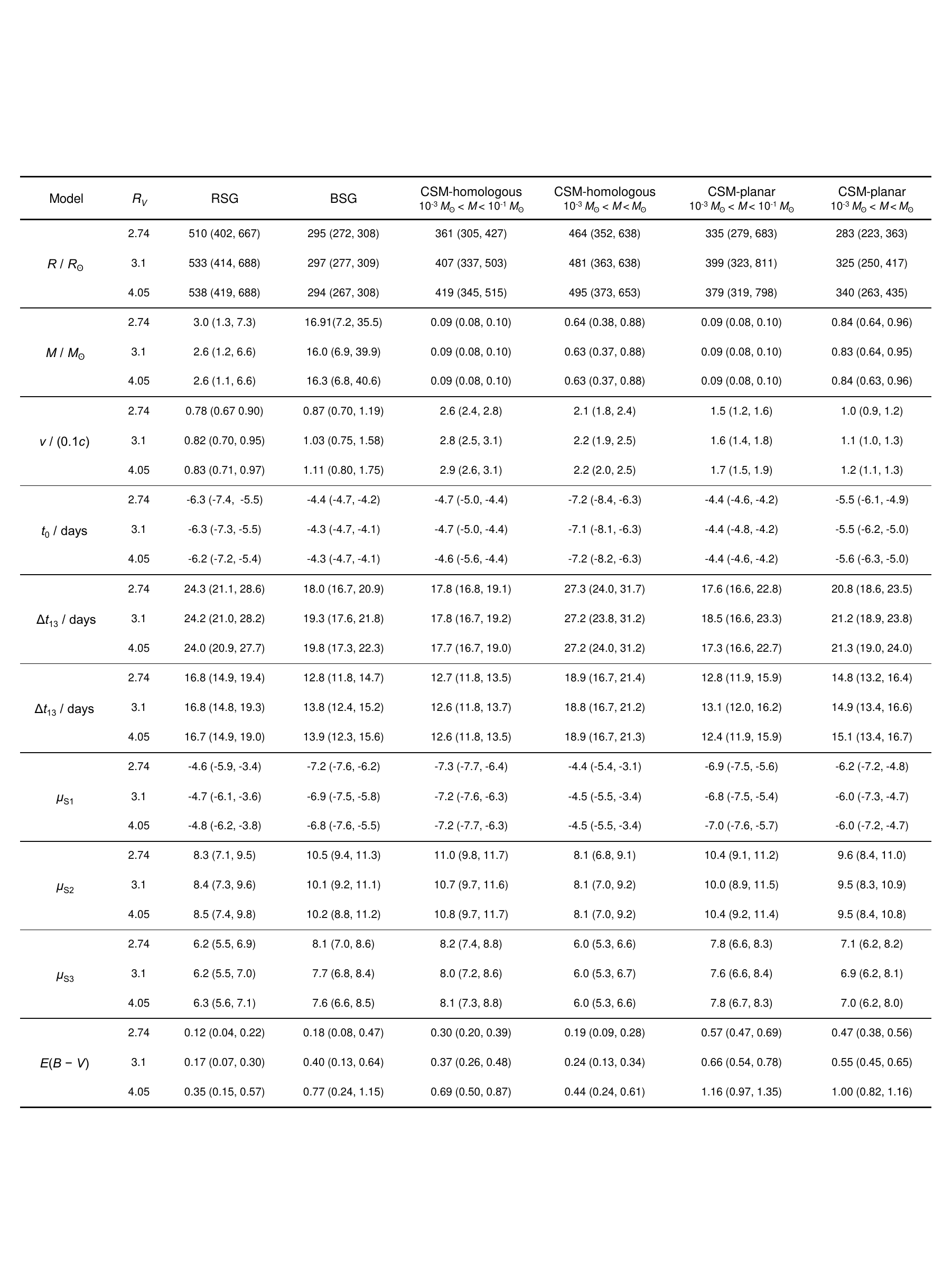}
\end{figure}
\clearpage
\thispagestyle{empty}
\subsection{Extended Data Table\,2}
Best-fit values of free model parameters from light curve models. Numbers in the parentheses are 68\% confidence intervals.


\end{document}